\documentclass[amsmath,nofootinbib,twocolumn,longbibliography,floatfix]{revtex4-2}

\pdfoutput=1 

\RequirePackage{graphicx} 
\usepackage{hyperref}
\usepackage{isotope}
\usepackage{float}

\newcommand{\expn}[1]{\ensuremath{\times 10^{#1}}}  
\newcommand{\refeq}[1]{Eq.~\ref{#1}}
\newcommand{\reffig}[1]{Fig.~\ref{#1}}
\newcommand{\refsec}[1]{Section~\ref{#1}}
\newcommand{\refapp}[1]{appendix~\ref{#1}}
\newcommand{\fspam}{F_{\text{BC}}}

\begin{document}

\title{Fast and high-fidelity state preparation and measurement in triple-quantum-dot spin qubits}
\date{\today}
\author{Jacob Z. Blumoff} \email[]{jzblumoff@hrl.com}
\author{Andrew S. Pan} 
\author{Tyler E. Keating} 
\author{Reed W. Andrews} 
\author{David W. Barnes} 
\author{Teresa L. Brecht} 
\author{Edward T. Croke} 
\author{Larken E. Euliss} 
\author{Jacob A. Fast} 
\author{Clayton A. C. Jackson} 
\author{Aaron M. Jones} 
\author{Joseph Kerckhoff} 
\author{Robert K. Lanza} 
\author{Kate Raach} 
\author{Bryan J. Thomas} 
\author{Roland Velunta} 
\author{Aaron J. Weinstein} 
\author{Thaddeus D. Ladd} 
\author{Kevin Eng} 
\author{Matthew G. Borselli}  
\author{Andrew T. Hunter} 
\author{Matthew T. Rakher} 

\affiliation{HRL Laboratories, LLC, 3011 Malibu Canyon Road, Malibu, California 90265, USA}

\begin{abstract}
We demonstrate rapid, high-fidelity state preparation and measurement in exchange-only Si/SiGe triple-quantum-dot qubits. 
Fast measurement integration (980 ns) and initialization ($\approx$300~ns) operations are performed with all-electrical, baseband control.   
We emphasize a leakage-sensitive joint initialization and measurement metric, developed in the context of exchange-only qubits but applicable more broadly, and report an infidelity of $2.5\pm0.5\expn{-3}$.  
This result is enabled by a high-valley-splitting heterostructure, initialization at the 2-to-3 electron charge boundary, and careful assessment and mitigation of $T_1$ during spin-to-charge conversion.  
The ultimate fidelity is limited by a number of comparably-important factors, and we identify clear paths towards further improved fidelity and speed. 
Along with an observed single-qubit randomized benchmarking error rate of 1.7$\times 10^{-3}$, this work demonstrates initialization, control, and measurement of Si/SiGe triple-dot qubits at fidelities and durations which are promising for scalable quantum information processing. 
\end{abstract}

\maketitle

\section{Introduction}
Spins in semiconductor quantum dots are promising foundational elements for scalable quantum information processing \cite{loss1998,vanderwiel2002,hanson2007,ladd2010,zwanenburg2013,ladd2018}.	
This technology leverages the hard-earned knowledge and exquisite tools of the semiconductor industry, which routinely mass-produces devices with billions of nanoscale structures.  
In this work we employ quantum dots which are laterally confined by lithographically-patterned gate electrodes and vertically confined within a high-mobility Si/SiGe heterostructure~\cite{eriksson2004,maune2012}.  
This construction yields suppressed sensitivity to gate-oxide disorder and noise relative to MOS-based devices~\cite{reed2016, kawakami2016, connors2019}.
Additionally, the use of isotopically-enriched \isotope[28]{Si} permits long spin-coherence times relative to control times~\cite{tyryshkin2012,eng2015,struck2020}.
Rather than form a qubit from an individual electron spin, one can encode a qubit in the collective spin state of three electrons in three quantum dots---the exchange-only qubit (more specifically, a decoherence-free subsystem [DFS] qubit)~\cite{lidar1998, divincenzo2000, kempe2001,fong2011}.  
These qubits are insensitive to global magnetic noise processes and allow universal control with only baseband electrical pulses, not requiring lasers, large RF circuits, or micromagnets. 

High-fidelity single-qubit operations for Si/SiGe exchange-only qubits have been previously demonstrated~\cite{andrews2019}, and we here show high-fidelity state preparation and measurement (SPAM) in a similar device.  
State preparation and measurement are integral to quantum computing---indeed they are the first and last steps in any algorithm, be it for NISQ, fault-tolerant computing, networking, or sensing applications.  
A reasonable ambition for SPAM performance is that it maintain fidelities and durations comparable to those of coherent manipulations, minimizing its likelihood of being a limiting factor in a larger application which may require multiple rounds of initialization and measurement.  
This establishes concrete goals and forms a natural basis of comparison, and using these criteria we show that exchange-only-qubit SPAM can maintain performance parity. 

Measurement of spin qubits has a deep history and admits a large space of approaches \cite{vanderwiel2002,hanson2007,zwanenburg2013}.
With few exceptions, most schemes involve a spin-to-charge conversion (S2C) mechanism followed by a measurement of the resulting charge-occupancy state.
Spin-to-charge conversion is most frequently achieved with spin-dependent tunneling (SDT)~\cite{elzerman2004b} or with Pauli spin blockade (PSB) \cite{ono2002,johnson2005}.
There are myriad quantum-dot charge detection technologies, foremost the use of quantum point contacts (QPCs) \cite{field1993,elzerman2004b}, single-electron transistors (SETs) \cite{kane2004,podd2010,morello2010,dehollain2014,watson2015,broome2017,keith2019,chanrion2020,seedhouse2021}, or dot charge sensors (DCSs, also called sensor quantum dots) \cite{hofmann1995,onac2006,nordberg2009}, all of which can be measured at DC or RF \cite{reilly2007,barthel2010,schoelkopf1998}.  
In the RF case, gate reflectometry can enable measurement at almost any gate electrode, and has yielded remarkable results~\cite{petersson2010,petersson2012,stehlik2015}.
This approach bolsters applicability to dense gate architectures~\cite{crippa2019,urdampilleta2019,ciriano-tejel2021,ansaloni2020,chanrion2020} but requires a bulky $LC$ resonator elsewhere in the system.
Here we employ PSB, which permits higher measurement fidelity and greater flexibility in applied magnetic field than does SDT, and a DCS, which requires only a structure of similar scale to---and simultaneously fabricated with---the qubits themselves.
Furthermore, DCSs allow for greater measurement sensitivity than do QPCs \cite{barthel2010}. 

Our key advancements relative to prior work in PSB/DCS spin qubit measurements---enabling the SPAM performance reported here---include details of the device, the control and amplifier chains, and our voltage biasing protocols.  
We also employ careful metrics for device-relevant characterization of SPAM fidelity, and much of this analysis is applicable to measurement modalities other than the PSB/DCS approach.  
One goal of this work is to provide an updated basis upon which to evaluate the potential of the PSB/DCS approach; whether it is sufficient for high-quality SPAM, or whether more read-out resources will be required for acceptable performance.

Spin qubit initialization also has several approaches, including the use of microwave pumping \cite{friesen2004}, relaxation \cite{elzerman2004b, petta2005}, and measurement-based mechanisms \cite{nakajima2019}, but in light of the desirably-weak interaction of spins with electromagnetic fields, these approaches are often impractically slow.    
A faster method available in spin systems is to exchange quantum-dot electrons with those from a cold electron reservoir.  
This contrasts with initialization in other architectures (e.g., superconducting qubits or trapped ion qubits), for which it is impossible or inefficient to deterministically replace the particles hosting the qubit.

This article is organized into the following sections: first we describe the measurement approach used for our devices and quantify some aspects of its performance, paying attention to both fidelity and duration. 
We then do the same for state preparation. 
We conclude by discussing our preferred technique for quantifying SPAM performance using a joint metric ($\fspam$, with $BC$ standing for Benchmarking Contrast) and compare the results to other key qubit device measures.  
\section{Measurement}
\begin{figure}
	\includegraphics[width=1\linewidth]{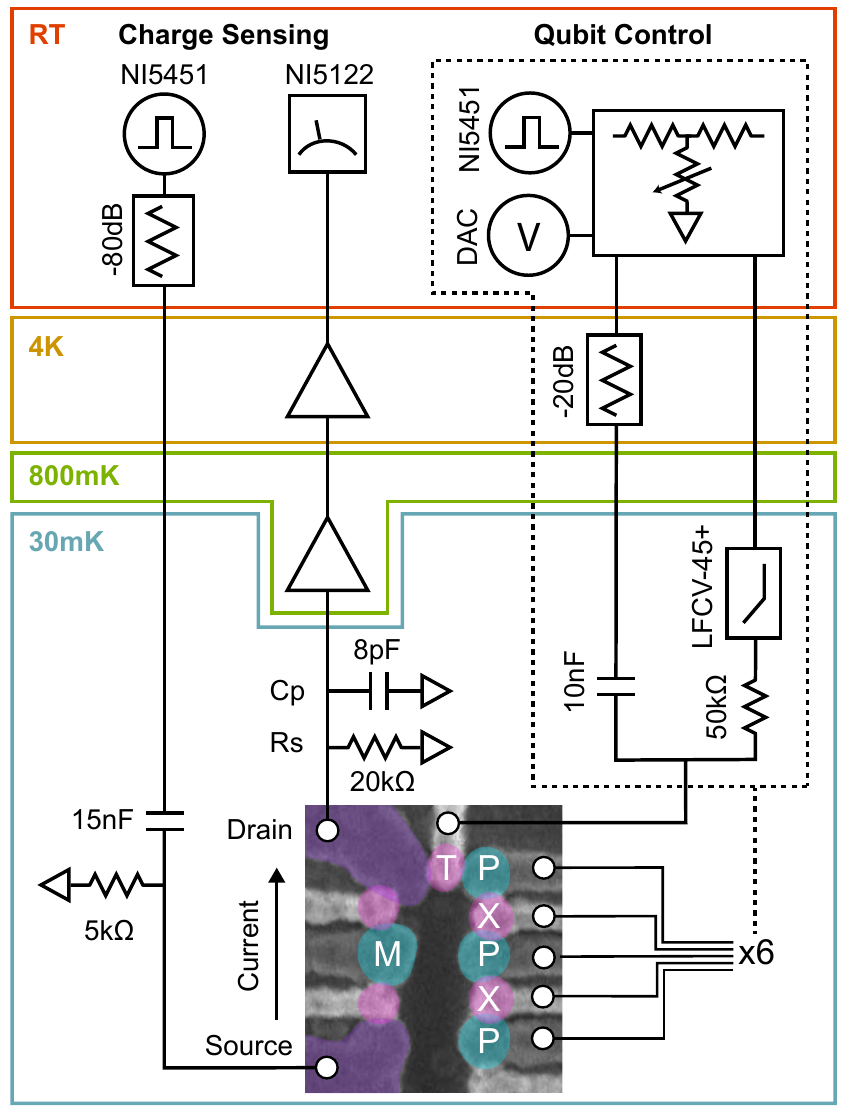} 
	\caption{
		\textbf{The signal chain.} Square waves are generated at room temperature, attenuated, and sent into the dilution refrigerator, where they act as a bias between source and drain two-dimensional electron gas (2DEG) leads.  
		These square waves probe the conductance of the DCS (labeled M), which is sensitive to the TQD charge state, and generate a current across a subsequent 20 k$\Omega$ sense resistor.  
		This signal is amplified by a two-stage cryogenic HEMT, then demodulated and thresholded at room temperature.  
		Signal adder circuits (dashed contour) enable fast and accurate baseband control of the plunger (blue, ``P'' \& ``M'') and barrier (pink ``T'' and ``X'') gate biases.  
		The adder circuit (simplified in this figure) is described in further detail in \refapp{section:app-hsa}.
		Note that the source signal attenuation has a low-impedance output.
	}
	\label{fig:signal_chain}
\end{figure}

In this work we employ a device fabricated similarly to that of Ref.~\onlinecite{andrews2019}.
This device comprises a linear array of six quantum dots and two DCSs, which share a common source but have independent drain leads. 
Only half of the device is utilized in this work, and a nominally identical device is shown in the electron microscopy inset of \reffig{fig:signal_chain}.
The overlapping gates are formed from aluminum, 60 nm above a 3-nm-wide, 800-ppm \isotope[29]{Si} quantum well~\cite{Zajac2015,Borselli2015,eng2015}.  

Singlet-triplet measurement via Pauli spin blockade exploits the fermionic nature of electrons, which are constrained to anti-symmetric wavefunctions.
At low magnetic fields, the ground state of two electrons in one dot must correspond to an anti-symmetric spin singlet; higher-energy triplet states which are spin-symmetric require valley and/or orbital excitations to provide overall antisymmetry.  
Here we notate charge in a set of dots as $(n,m)$; for example (2,0) represents two electrons in dot 1, zero electrons in dot 2, etc.  
A spin singlet is indicated by S, so ``S(2,0)" is the ground state of the (2,0) charge state of a double-dot.  
In the (2,0) charge state, valley and/or orbital excitations provide a singlet-triplet energy splitting $\delta_{ST}$ which can be exploited for both spin initialization and measurement, with the magnitude of that splitting being critical to the fidelity of both operations.

In a triple-quantum-dot (TQD) DFS, the encoded-$|0\rangle$ state is given by a spin singlet on two (hereafter ``outer'') dots, ``S(1,1)''.
The third dot charge-state is omitted for brevity but is singly occupied by an electron which is sometimes referred to as the ``gauge spin'' and is crucial for exchange-only universal control. 
This electron need not be initialized into any particular spin state~\cite{andrews2019,fong2011}, but reliable exchange coupling requires that it not be in an excited orbital or valley state.  
The encoded-$|1\rangle$ state of the DFS qubit is a superposition of triplet-states [T$_{0,\pm}$(1,1)] on the outer dot spins, entangled with this third spin.
Since PSB measures whether or not the two outer spins are in a spin singlet or triplet, the preparation and measurements of exchange-only qubits can proceed in the same fashion as in double-quantum-dot singlet-triplet qubits~\cite{levy2002,barthel2009}.

Charge-states of the triple-quantum-dot are probed by the electrostatic interaction with a nearby DCS (underneath the ``M'' gate in the device depicted in \reffig{fig:signal_chain}).  
The DCS is biased into the Coulomb blockade regime, where its conductance is highly sensitive to the potential environment and, accordingly, to the charge occupancy of the TQD.
We interrogate this conductance by applying a bias between the source and drain 2DEGs connected to this dot, modulated by a square wave at a frequency of about 2 MHz.
The resulting DCS drain current flows through a 20 k$\Omega$ sense resistor $R_s$, and the generated voltage signal is amplified by a two-stage cryogenic HEMT (Avago ATF-38133) \cite{tracy2016,vink2004}, shown in \reffig{fig:signal_chain}.
Placing a low-dissipation stage at the still plate ($T\approx800$ mK) before  a second stage of amplification (at $T\approx4$ K) allows for reduced added noise while respecting the thermal constraints of the dilution refrigerator.
A cold finger locates that 800 mK amplifier in spatial proximity (several cm) to the device chip, minimizing the (bandwidth-limiting) parasitic capacitance $C_p$, which we estimate to be roughly 8 pF from the signal bandwidth.
The signal then continues to room temperature where it is digitized and demodulated in software.

\begin{figure}
\includegraphics[width=1\linewidth]{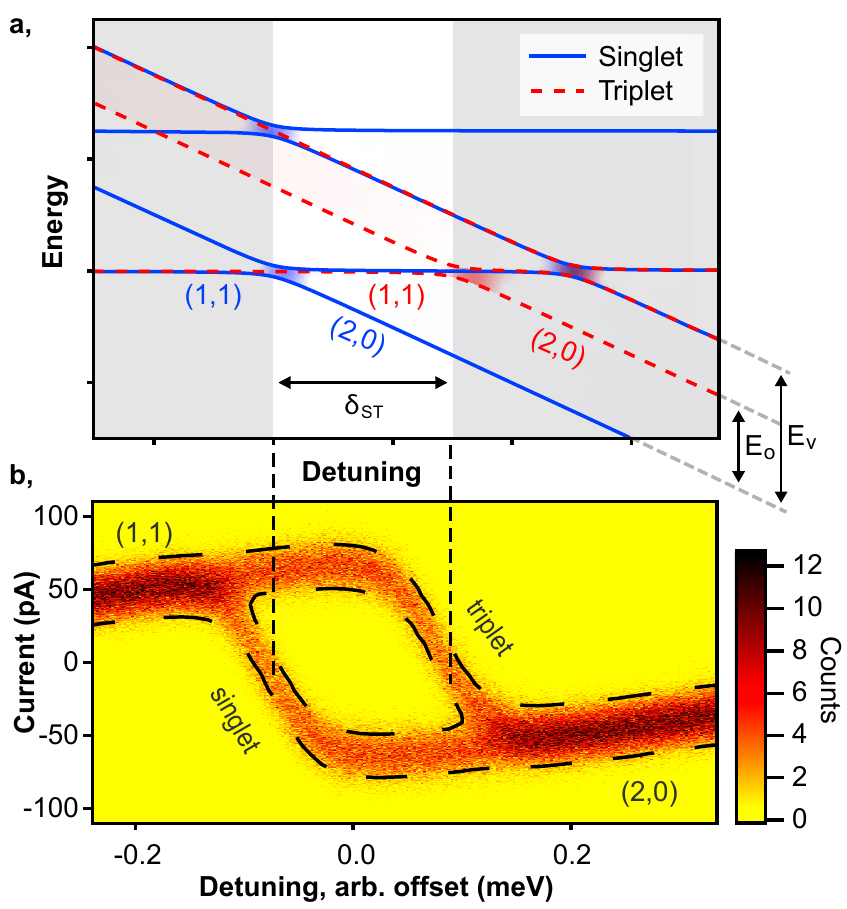} 
\caption{
	\textbf{Energy level diagram and spin blockade spectroscopy. }
	\textbf{a,} Energy level diagram near the measurement window, shown at zero magnetic field for clarity.  
	Horizontal lines (energy independent of detuning) represent the (1,1) charge state used for exchange and idle operations, while the diagonal lines represent the (2,0) charge state.  
	In this illustration the relevant singlet-triplet splitting $\delta_{ST}$ (and accordingly the region of spin-to-charge conversion) is set by the two-electron orbital energy ($E_o$) rather than the excited valley state energy ($E_v$).
	\textbf{b,} Spin blockade spectroscopy, showing histograms of DCS current as a function of detuning between the (2,0) and (1,1) charge states.  
	The fully-resolved gap between the singlet and triplet branches is reflective of highly-efficient spin-to-charge conversion. 
}
\label{fig:spectrum_and_sbs} 
\end{figure}

We measure the spin state of the device by performing spin-to-charge conversion---detuning the device from an ``idle'' bias configuration (or ``coordinate'') of low exchange energy in the (1,1) charge cell towards the (2,0) regime.
Because of PSB and the finite singlet-triplet energy splitting, the charge-state level crossing occurs at higher detuning bias for spin triplets than for singlets (see the energy diagram in \reffig{fig:spectrum_and_sbs}a.)
This sets a ``measure window'' of bias where the spin states differ in their charge character.
At an intermediate bias near one half of the two-electron excited-state splitting, the singlet-triplet degree of freedom and the charge occupancy are maximally correlated, and measurement of charge also yields spin-state information. 
This is directly evident from spin blockade spectroscopy---histograms of DCS conductance as a function of detuning bias (\reffig{fig:spectrum_and_sbs}b) \cite{jones2019}.

Near the charge-state transition the bias is ramped in time so as to minimize Landau-Zener processes which might excite the spins into elevated valley-orbit states.
This is discussed further in \refsec{section:init}. 
Due to imperfect signal integrity engineering, with the current design we require a ``settling'' period of order one to ten $\mu$s before the source-drain square waves are applied to maintain consistent measurement results.
Generally this delay, rather than integration, dominates the obtainable measurement cadence, but we expect to ameliorate this with further signal chain engineering.

\subsection{SNR} \label{section:results_snr}
\begin{figure}
\includegraphics[width=1\linewidth]{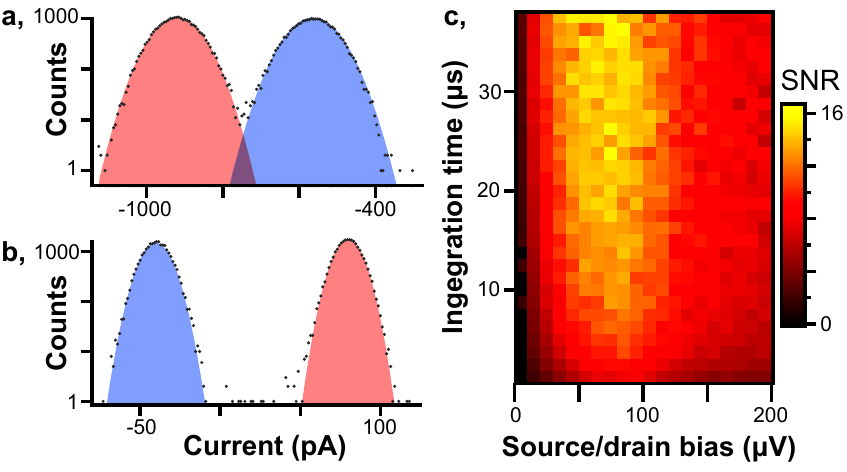} 
\caption{
	\textbf{Readout histograms and SNR as a function of time and bias amplitude. } 
	\textbf{a,} Log-scale histograms of DCS current (data in black), with Gaussian fits indicating singlet (blue) and triplet (red) readout signals, here with a total integration time of 980~ns.  
	Thresholding a distribution with this SNR (6.5) provides a bound on $1-\fspam \geq 6\expn{-4}$. 
	\textbf{b,} Histograms taken at an integration time of 40 $\mu$s and with an additional measurement subtracted, giving an SNR of 15.5.  
	This distribution bounds $1-\fspam \gtrsim 5\expn{-15}$, but other limiting effects (such as relaxation during measurement) are evident.  
	\textbf{c,} Histogram SNR as a function of both source-drain bias amplitude and integration time.  
	SNR saturates as a function of increasing integration time due to $1/f$ noise on the DCS potential.  
	SNR increases with bias amplitude, then decreases after a critical point.
}
\label{fig:snr_and_hist}
\end{figure}

We now probe the quality of that measurement, starting with the signal to noise ratio (SNR).
The experimental routine begins with preparation of a dephased spin state, which is created by waiting at the idle bias for a time longer than the hyperfine-dominated $T_2^*$.
This is followed by the previously-described S2C and charge-state measurement processes, and we histogram the resulting measurement current.
We use this routine to explore the obtainable spin SNR as a function of measurement duration and strength (i.e. source-drain bias amplitude) in \reffig{fig:snr_and_hist}.
As we increase the integration duration from 980~ns to roughly 40~$\mu$s, the SNR first improves as we better average away white noise, then saturates as it becomes limited by the $1/f$ charge noise on the DCS.
Similarly, the SNR increases quickly as we increase the source-drain bias amplitude but then decreases after a critical value.
This reflects nonlinearity in the DCS conductance, for instance due to excited states.
Quantitative predictions for both components of SNR are discussed in detail in \refapp{section:app-SNR}.

We can further look at two extremal cases:  the shortest-duration measurement which maintains a ``sufficiently high'' SNR, and the long-duration measurement with maximized SNR.
\reffig{fig:snr_and_hist}a depicts a rapid measurement, integrating for only 980~ns.
This timescale compares favorably to spin-relaxation $T_1$ ($>$10~ms, discussed further in \refsec{section:disc_t1}), quantum-dot dynamical-decoupling timescales \cite{veldhorst2014}, and anticipated two-qubit gate durations \cite{andrews2019, fong2011}.
In that time we obtain an SNR of 6.5, which limits\footnote{When noise on the singlet and triplet readout signals is Gaussian and symmetric, SNR bounds fidelity by the relation
	\begin{equation}
		1-\fspam\geq\frac{1}{2}\left(1-\text{erf}\left(\frac{\text{SNR}}{2 \sqrt{2}}\right)\right).
\end{equation}} $1-\fspam$ to 6\expn{-4} ($\fspam$ is defined carefully in \refsec{section:joint}).

Alternately we can saturate the high-SNR limit by integrating for 40~$\mu$s 
(and subtracting a reference measurement), as shown in \reffig{fig:snr_and_hist}b, which achieves an SNR of 15.5.
This SNR yields an exceptionally small theoretical bound on infidelity, but other factors limit fidelity far more strongly.
An examination of the histogram of \reffig{fig:snr_and_hist}b shows evidence of $T_1$ (points scattered between the two Gaussian distributions) and additional spurious counts, the latter of which may indicate more-pathological noise mechanisms or a probability of populating other charge states.  
We now go on to discuss those $T_1$ effects.

\subsection{Relaxation during measurement}\label{section:disc_t1}
\begin{figure}
\includegraphics[width=1\linewidth]{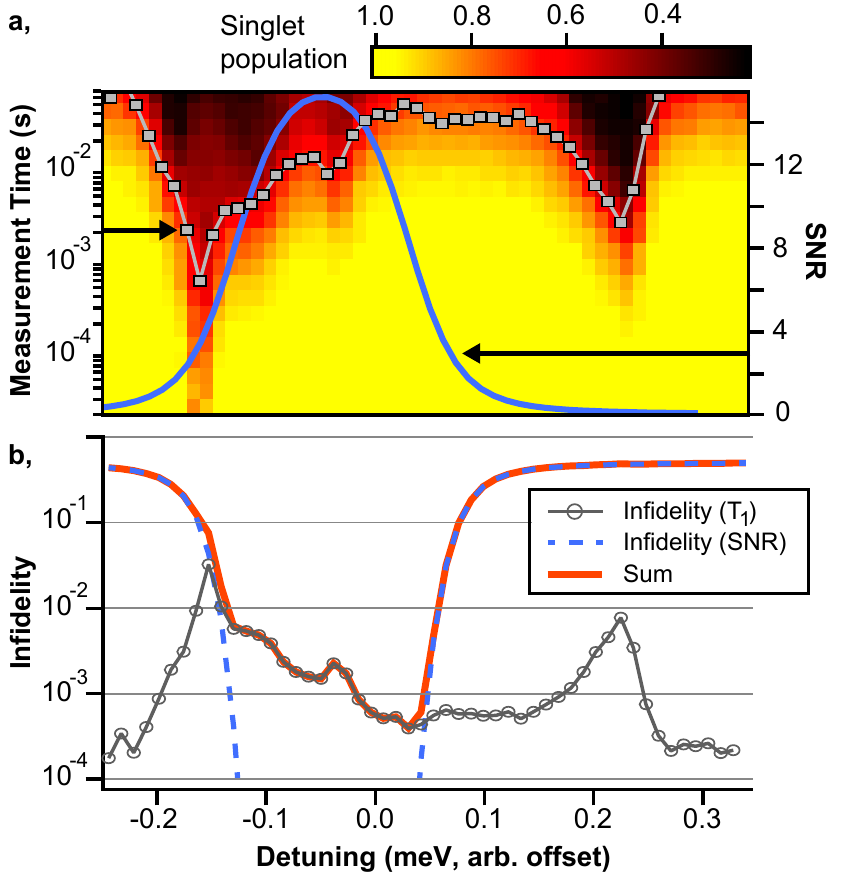} 
\caption{
	\textbf{$T_1$ and SNR as a function of measurement bias and the resulting limits on SPAM fidelity. }
	\textbf{a,} In color, the resulting singlet probability after trial measurements at swept bias (x axis) and duration (left vertical axis).  
	In gray, the fit $1/e$ times at each bias point.  
	In blue, the SNR as a function of bias, extrapolated from a fit of spin blockade spectroscopy (right vertical axis).   
	\textbf{b,} Calculating the limits on $\fspam$ that result both from finite SNR and from relaxation during measurement.  
	Combining those sources, we see clearly that the maximum fidelity measurement bias does not coincide with maximal SNR for this choice of measurement parameters.
}
\label{fig:t1_snr_vs_detuning}
\end{figure}

As we access larger signal-to-noise ratios, measurement fidelity quickly becomes limited by relaxation during the measurement. 
This was explored experimentally, and a key result is shown in \reffig{fig:t1_snr_vs_detuning}a.
These data were gathered by preparing a dephased qubit state and transitioning to a trial measurement bias, then sweeping the detuning of that bias.
For each detuning we apply a typical modulated source-drain bias for a variable amount of time (called ``measurement time'' in the figure), but do not record a measurement signal.
After that time we perform a calibrated measurement at a fixed bias to examine the resulting state, and we show that result as a function of both trial measurement detuning and duration.
The detuning axis has an arbitrary offset but is referenced to the SNR at each point (extracted from spin blockade spectroscopy), which serves to demarcate the singlet and triplet charge-state transitions.

The locations of the sharp reductions in $T_1$ at negative detuning are consistent with anticrossings between the S(1,1) and T$_-$(1,1) states, given our 1.5~mT applied magnetic field.
We can estimate the bound on measurement fidelity imposed by finite SNR and $T_1$, as shown in \reffig{fig:t1_snr_vs_detuning}b\footnote{Relaxation during measurement yields a bound
\begin{equation}1-\fspam \geq 
\frac{1}{2} 
\left(1-\exp{\left(-\frac{T_{\text{measurement}}}{T_1}\right)}\right)
\end{equation}}. 
Combining those two effects, we can locate the bias for optimal contrast---which, notably, is not the bias at which SNR is maximized.
Importantly, we see $T_1$ decay times of over 20~ms within biases that yield high SNR, which compares favorably to our achievable measurement rates of $\mu$s order.  
We additionally performed the same experimental sequence and analysis but with the amplitude of the source-drain bias as the independent variable, and discuss that result in \refapp{section:app-t1sd}.
That measurement shows that large source-drain bias can induce $T_1$, and accordingly a compromise must be made in choosing that amplitude.

The rich $T_1$ behavior displayed in \reffig{fig:t1_snr_vs_detuning}a involves a complex interplay of microscopic semiconductor physics. 
One extrinsic relaxation pathway is via cotunneling with the electron bath, but this can typically be exponentially suppressed by increasing the relevant tunnel barrier.
The intrinsic decay channels within the device require both charge-transition and spin-flip mechanisms. 
Charge decay proceeds via electron-phonon coupling or electromagnetic interactions with the gate, bath, or charge noise sources, which can have different densities of states and hence differing dependence on detuning.
Microscopically, singlet and triplet states can couple to each other via nuclear hyperfine interaction with \isotope[29]{Si} and \isotope[73]{Ge} atoms, as well as the spin-orbit interaction present in Si/SiGe quantum wells.
$T_1$ ``hot spots'' can appear at detuning biases where excited singlet and triplet states anticross, leading to strong spin mixing and rapid charge decay, as observed experimentally in \reffig{fig:t1_snr_vs_detuning}a. 
In \refapp{section:app-measwindow} we address how these features are expected to connect with the ``measure window'' energy spectrum.

\section{State Preparation}\label{section:init}
\begin{figure}
	\includegraphics[width=1\linewidth]{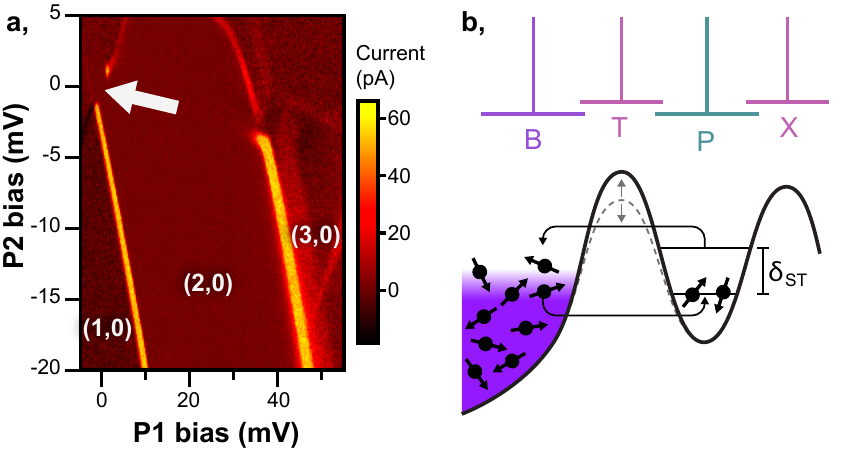} 
	\caption{
		\textbf{Sweeping initialization bias and initialization schematic.}
		\textbf{a,} This signal (color) is the difference of spin measurements between preparations of an initialized state and of a dephased state.  
		When initialization is effective, these measurement results are not identical.  
		Sweeping the plunger-gate initialization bias of the outer two dots, we see two bands at the (1,0)-(2,0) and (2,0)-(3,0) charge cell boundaries, which are both usable regions for initialization.  
		The light gray arrow indicates the bias region used for S2C during measurement.   
		\textbf{b,} Near the (2,0)-(1,0) charge state transition the tunneling rate between the outermost dot and the 2DEG bath is greatly accelerated, leading to rapid thermalization and depopulation of the (2,0) excited states.  
		The detuning span of this region is set by the two-electron singlet-triplet splitting.
		The rate can be further increased by dynamically depressing the dot-bath barrier (dashed line), physically realized as the ``T'' gate visible in \reffig{fig:signal_chain}.
	}
	\label{fig:doflush_and_schematic} 
\end{figure}

We initialize the spin state  by biasing fully into the (2,0) charge configuration, maximizing the energy splitting between the triplet states and the ground-state singlet.
At thermal equilibrium, we expect to preferentially populate the encoded-$|0\rangle$ (singlet) state. 
Waiting for relaxation to equilibrium is impractical, as singlet-triplet decay rates ($1/T_1$) can be quite long due to the necessity for spin relaxation \cite{hayes2009,Prance2012}.
Instead, as previously mentioned, we speed up the process by exchanging electrons with a cold reservoir---here a 2DEG which also serves as the drain for the DCS current.
Typically initialization is achieved by biasing the outermost dot to the vicinity of the (2,0)-(1,0) charge boundary where the chemical potential of the (1,0) charge state lies between that of the S(2,0) and T(2,0) states (as shown empirically in \reffig{fig:doflush_and_schematic}a and schematically in \reffig{fig:doflush_and_schematic}b).
This enables bath-dot tunneling to more rapidly equilibrate or ``flush'' to the ground state.
We sweep the duration of this flush period and choose a duration at which the triplet population stabilizes.
In this work, we exploit the (2,0)-(3,0) charge boundary [rather than (1,0)-(2,0)], seen on the right side of \reffig{fig:doflush_and_schematic}b, which to our knowledge has not been previously discussed for this purpose.

Initialization is completed by transitioning back to the idle bias via a ramped trajectory.
First, we change the bias quickly from the initialization charge boundary to an ``entry'' point just outside the (2,0)-(1,1) boundary.
Then we ramp slowly---with ramping times of order 10-100~ns---into and through the PSB coordinate, ending at a second entry point just inside the (1,1) charge cell.
This is designed so as to move through the anti-crossing without triggering a Landau-Zener transition. 
Then, we jump quickly from that point to idle, minimizing the time spent at low-exchange regions where magnetic dephasing is more rapid. 
To return from idle to measurement, we repeat latter half of this process in reverse. 
The optimal choices of ramp times and the entry coordinates balance many factors, and are a subject for future study.

\subsection{Initialization fidelity} \label{section:init_spec}
After tunneling to and from the bath, at equilibrium we expect to populate the ground-state singlet state with probability given by the partition function
\begin{equation} \label{eqn:partition_function}
	P_{|0\rangle}^{-1} \approx Z_{2e} \approx 1+3e^{-E_o \beta}+4e^{-E_v \beta}\approx\left(5\times 10^{-4}\right)^{-1},
\end{equation}
where $\beta=1/(k_B T_e)$ (here, the effective electron temperature $T_e\approx 220$~mK, measured from tunneling linewidth as a function of mixing chamber temperature), $E_v$ is the excited valley state energy in the outermost dot ($\approx$250~$\mu$eV, measured with detuning axis pulsed spectroscopy \cite{chen2021}), $E_o$ is the two-electron orbital excited state energy ($\approx$160~$\mu$eV, measured with spin blockade spectroscopy), and we neglect higher excited valley-orbital states (assuming their contributions are negligible at sufficiently low temperatures).
The factors of three and hour come from the respective degeneracies of the excited valley and orbital states.
Zeeman splittings of the polarized triplet states are also neglected in this  equation as they are small at the low and moderate magnetic fields at which we typically operate exchange-only qubits (here, 1.5 mT).
We have also neglected the third electron, which we estimate to be separable from the outer-dot initialization, and may eventually reach an equilibrium set by its own (one-electron) excited-state energy.
It is possible to end in the wrong charge state during this process, so for example the partition function in \refeq{eqn:partition_function} can be extended to account for (1,0) occupation as well (if initializing at that charge boundary), though empirically this is not a common outcome.

\subsection{Initialization speed} \label{section:init_speed}
\begin{figure}
	\includegraphics[width=1\linewidth]{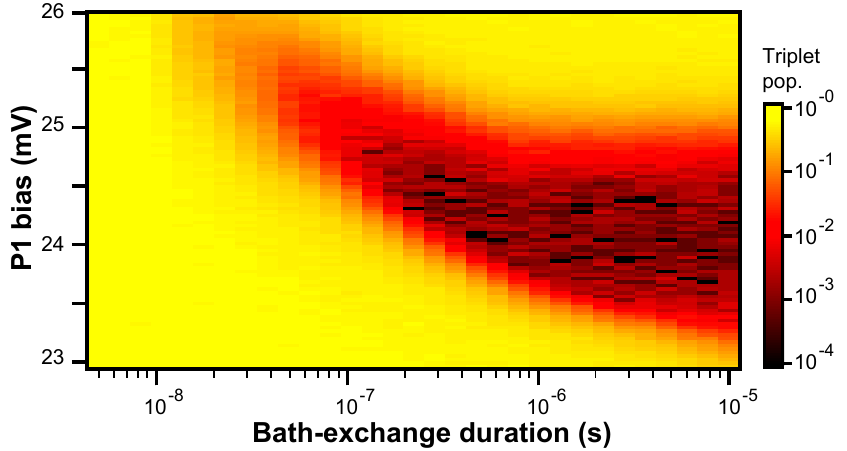} 
	\caption{
		\textbf{Sweeping initialization bias and duration.}
		Sweeping the outer-dot plunger gate bias across the (2,0)-(3,0) boundary and sweeping time.   
		This population is the thresholded result of a spin-state measurement immediately after bath-exchange following preparation of a dephased state.  
		The populations are not monotonic in duration, which is indicative of the bias waveforms settling in time.  
		While our high-speed adder circuit (\refapp{section:app-hsa}) attempts to provide a flat response, not all non-idealities are accounted for. 
		This settling limits the minimum initialization time in this case to roughly 300 ns to reach an asymptotic population of 6\expn{-4}.  
		P1 biases differ from \reffig{fig:doflush_and_schematic}a because additional (pulsed) tunnel-barrier bias has been introduced and that T gate has a non-trivial cross-capacitance to the first dot potential.
		This result is degraded from the Boltzmann distribution prediction for initialization fidelity alone (\refeq{eqn:partition_function}, $\approx$5\expn{-4}), due to measurement and mapping errors.  
	}
	\label{fig:swoosh}
\end{figure}

This speed of this initialization process benefits from the fact that tunneling between the outer dot and bath depends exponentially on the height of the insulating tunnel barrier.
Pulsing the voltage bias on the relevant barrier gate strongly and quickly modulates that potential.
In practice, we observe a speed limit enforced by two main mechanisms:  orbital suppression and waveform stability.
First, depressing the barrier height can distort and desymmetrize the quantum dot confining potential, which decreases the orbital excited-state energy \cite{melnikov2006} and limits the ultimate obtainable fidelity (\refeq{eqn:partition_function}). 
In principle, this effect could be countered by shaping the potential further with appropriate pulses on other nearby gate electrodes.
Second, the stability of the bias waveform strongly influences this process.
Ideally, when sweeping the initialization bias and duration (\reffig{fig:swoosh}), we expect the resulting population to be strongly bias-dependent and to change monotonically with time.
In contrast, we believe the observed non-monotonic signal is a signature of bias drift.
This effect limits us to a minimum duration of roughly 300~ns to reach the asymptotic population; however this limitation should be reduced with improved engineering for signal integrity.
We note also that a wider initialization bias window (resulting from larger excited-state splittings) should also reduce the sensitivity to this drift.

We find that using the alternate (2,0)-(3,0) ``initialization window'' leads to the same asymptotic singlet population but often more than an order of magnitude more quickly.
Several factors likely contribute to this operational improvement; for one, the increased dot plunger (P) gate voltage required at this boundary decreases the effective tunnel barrier.
Second, the increased bias-space width of this initialization window (evident in \reffig{fig:doflush_and_schematic}) likely mitigates the effects of waveform drift.
Finally, the spin and charge character of the multi-electron states also play an important role. 
For instance, it can be shown from the relevant Clebsch-Gordan coefficients that the tunneling transition rate between an $N$-electron state of total spin $S_i$ and an $N+1$ state with spin $S_j$ is weighted by a factor \cite{weinmann1994}
\begin{equation}\label{eq:spin_dep_tunnel_rates}
	\gamma_{ij}=\frac{S_i+1}{2S_i+1} \delta_{S_i+1/2,S_j }+\frac{S_i}{2S_i+1} \delta_{S_i-1/2,S_j }.
\end{equation} 
As a result, while both the one- and three-electron ground states have total spin $S$ = 1/2 at typical magnetic fields, the single-electron state has a three-times larger tunneling rate to the triplets compared to singlets where both are energetically available, whereas the opposite is true for the three-electron ground state, which has a three-times larger tunneling rate for singlets compared to triplets.
Details of the three-electron excitation spectrum may also play a role and depend sensitively on the confining electrostatics as well as valley mixing; quantifying the effects of these states on the initialization process is an interesting area for further study.

\section{Joint SPAM fidelity \& budget} \label{section:joint}
\subsection{Quantifying SPAM performance}
In this section we move on from the mechanics of individual SPAM operations to focus on the quantification of their performance.  
Measurement SNR and $T_1$ both contribute to SPAM infidelity and are useful diagnostic quantities, but tell only part of the story. 
We can probe initialization fidelity by performing a measurement directly after initialization (as in \reffig{fig:swoosh}) however we do not consider this quantity to be a reliable summary metric of initialization quality, nor of integrated SPAM quality.  
Among other concerns, it can be gamed by biased measurement error, it doesn't include any errors in ``mapping'' when biasing to and from the idle point, and it does not guarantee what fraction of the population will exhibit proper qubit evolution.
The last criterion is generally relevant due to the possibility of leakage outside of the qubit space, the effects of which vary widely between physical-qubit implementations.
One plausible alternative metric is the ultimate contrast that can be observed in exchange oscillations (\refapp{section:app-exchange}), but this is also degraded by charge noise and other decoherence mechanisms whose infidelity is better associated with coherent manipulation.

Instead, we emphasize an integrated preparation and measurement metric (first presented in Ref.~\onlinecite{andrews2019} and reproduced mathematically in \refapp{section:app-fspam}) that avoids the previously mentioned flaws while providing several other benefits.
This measure, dubbed $\fspam$ for Benchmarking Contrast, probes the fundamental concern of imperfect preparation and measurement:  what fraction of the time do we obtain the correct result after an arbitrary qubit evolution, independent of the fidelity of that evolution.
$\fspam$ is derived from ``blind'' randomized benchmarking, a method originally designed to detect leakage out of the computational space during coherent manipulation of exchange-only qubits.
In this procedure an ordinary randomized benchmarking experiment (with sequences that compile to the identity) is executed along with a second benchmarking experiment, whose sequences are instead engineered to compile to a population-inverting gate.
To assess SPAM performance, we evaluate the contrast of these fit curves at zero Clifford operations, which attempts to subtract any contrast loss due to imperfect qubit rotations.

$\fspam$ transposes randomized benchmarking's typical objective--measuring Clifford fidelity independently of SPAM performance.
This measure differs from assignment fidelity\footnote{$F_A=1-\frac{1}{2}\left(P[0,1]+P[1,0]\right)$, where the latter terms indicate the probability of registering an erroneous binarized measurement result after nominal preparation of a computational state.}, which might be quite coarsely approximated by the same curves evaluated at 0.5 Clifford operations.
Importantly for exchange-only qubits, we find $\fspam$ to be more robust than assignment fidelity in the presence of leakage, though that import depends on the qubit physical implementation.
$\fspam$ also implicitly disallows the use of different processes for preparation or measurement of $|0\rangle$ and $|1\rangle$ which may boost assignment fidelity at the expense of operational utility.
We note that this metric is essentially universal, and allows fair comparison between diverse physical implementations of qubits.
One final desirable feature of $\fspam$ as a ``summary'' metric is its resilience to manipulation, or being increased at the expense of the fidelity of other qubit operations.
Any tradeoffs between SPAM and single-qubit errors are readily apparent in this extraction.

\begin{figure}
	\includegraphics[width=1\linewidth]{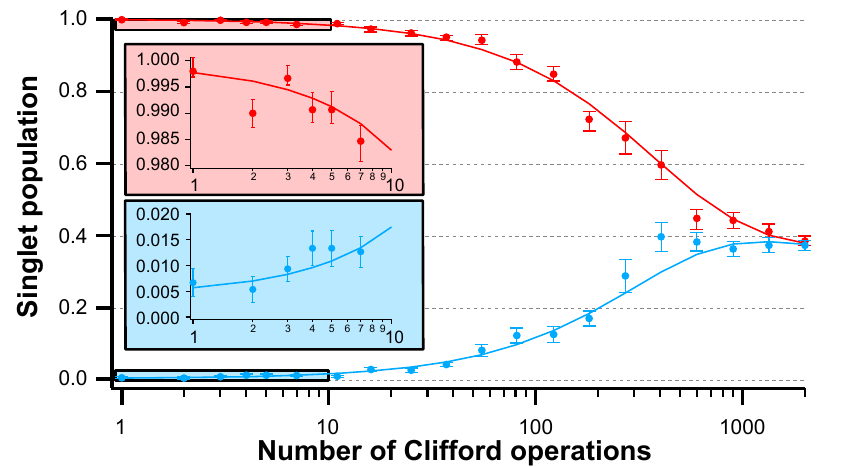} 
	\caption{
		\textbf{Measuring joint initialization and measurement fidelity from benchmarking contrast.} 
		The return-to-singlet (red) and return-to-triplet (blue) traces from blind randomized benchmarking.  
		The contrast of these fit curves, extrapolated to 0 Clifford operations,  indicates $1-\fspam = 2.5(5)\expn{-3}$, with a per-Clifford qubit error of $1.7(2)\expn{-3}$, and a per-Clifford leakage rate of $3.3(9)\expn{-4}$.
		Upper and lower insets emphasize full contrast at the shortest sequences.
	}	
	\label{fig:contrast_metrics}
\end{figure}

Blind randomized benchmarking for the device described in this work is shown in \reffig{fig:contrast_metrics}a.  These data indicate a per-Clifford error of 1.7\expn{-3} and $\fspam$ of 2.5(5)\expn{-3}.  This latter estimate is highly consistent with the visible contrast of exchange oscillations (\refapp{section:app-exchange}).  

\subsection{Compiled budget \& missing error}
To understand this SPAM performance, we can consider three components:  initialization, measurement, and mapping.
Initialization quality is foremost bound by the magnitude of the 2e and 1e excited state energies relative to the effective electron temperature $T_e$ (\refeq{eqn:partition_function}).
Measurement performance can be determined from a trio of ``active'' factors:  signal amplitude, noise amplitude, and relaxation during measurement---the latter of which was discussed in \refsec{section:disc_t1} and the former two in \refapp{section:app-SNR}.
These three factors are tightly interrelated and sometimes in tension, so that overall optimization is often a balancing act between them.
Mapping errors, which occur in the transitions from initialization to the idle bias and from the idle bias to the measurement bias, can reasonably be grouped with one of the above or considered as a separate category, but should be included in any operationally-representative integrated SPAM metric.
These errors are further discussed in \refapp{section:app-mapping}.
Here we consider and assemble the effects of those components into a budget for $\fspam$, shown in table \ref{tab:brb_budget}.

\begin{table}
\centering
\begin{tabular}{lc}
Factor & Error contribution \\ \hline
Boltzmann distribution & 5\expn{-4} \\
SNR (980 ns integration) & 6\expn{-4} \\
$T_1$ (measurement \& settle) & 2\expn{-4} \\
(1,1) mapping errors & 2\expn{-4} \\ \hline
Total & 1.5\expn{-3} 
\end{tabular}
\caption{Budget for $\fspam$.}
\label{tab:brb_budget}
\end{table}

This budget leaves roughly 1\expn{-3} infidelity unexplained relative to the observed $1-\fspam =2.5\expn{-3}$.
We hypothesize that the missing fidelity is attributable to probabilistically initializing with the gauge electron in an excited state.
During these events, the exchange rate on one control axis will exhibit a different voltage dependence ($J(V)$) than occurs when the gauge electron is in its ground state.
As a result, Clifford operations composed of pulses calibrated for the $J(V)$ mapping of the ground state quickly scramble the spin-state when that electron is excited, resulting in an apparent SPAM error.
This effect, sometimes described as ``dual exchange frequencies,'' has been  reported in other works~\cite{reed2016}, where the Fourier transform of time-domain exchange oscillations clearly shows two peaks.
For the present device under study, we are unable to observe a clear second peak in the spectra of exchange evolution on either exchange axis, however it is unclear whether this experiment has sufficient resolution to differentiate a second frequency at the 1\expn{-3} level. 
Development of a better method to quantify such population is an important area for future work.

Though the energy of the gauge electrons first excited-state was not measured, as a one-electron property we can reasonably associate its first excited state with the valley degree of freedom. 
Assuming equilibrium statistics in a Boltzmann distribution, the missing infidelity implies an excited valley-state energy of roughly 130 $\mu$eV, a value which is consistent with the measured valley splitting distribution for this heterostructure~\cite{chen2021}.
We note that this population would not be evident in initialization metrics which did not require fairly measuring qubit evolution, pointing again to the importance of the use of blind randomized benchmarking as a fair SPAM fidelity metric.

\section{Conclusion}
By our preferred, conservative metric, we have demonstrated preparation and measurement in exchange-only qubits with $1-\fspam$=2.5(5)\expn{-3}.
These are obtained using only baseband control electronics, using a measurement integration time of 980 ns, an initialization duration of roughly 300 ns, and a charge-transition ramp time of 100 ns.  
We present a budget which explains $\approx$60\% of the observed error, and hypothesize that the remaining error is due to thermal excited states of the gauge electron.  

Priorities for improving initialization performance are obtaining larger excited-state energies and suppressing the effective electron temperature, which should yield both improved singlet and gauge electron preparation.
Initialization may be accelerated by achieving more stable voltage waveforms.
Related signal integrity engineering is also key to reducing the ``settling'' period, which presently dominates the possible measurement cadence.
Other paths to improved PSB/DCS measurement performance include lower-noise amplification with lower parasitic capacitances, further exploration of ``latching''~\cite{studenikin2012,mason2015,harveycollard2018} and ``avalanche''~\cite{van2020} approaches, and improved $T_1$ decay rates.
$T_1$ effects can likely be reduced by further decreasing charge and magnetic noise, which are already desiderata for improved coherent manipulation.
This may also require using smaller source-drain biases, further emphasizing the need for improvements in signal and noise amplitudes.

Importantly, we conclude that high-fidelity and rapid SPAM (maintaining performance parity with coherent manipulation) is achievable in spin qubits using only simple electronics and components with high flexibility in temperature and magnetic field.  Routes to even further improvement without addition of other components are clear.

\section{Acknowledgments}
The authors gratefully acknowledge the important contributions of Andrey Kiselev, Matthew Reed, Laura de Lorenzo, Emily Pritchett, Cody Jones, and Ed Chen, and support in figure preparation from John Carpenter.

\bibliography{qdotref}

\begin{thebibliography}{73}%
\makeatletter
\providecommand \@ifxundefined [1]{%
 \@ifx{#1\undefined}
}%
\providecommand \@ifnum [1]{%
 \ifnum #1\expandafter \@firstoftwo
 \else \expandafter \@secondoftwo
 \fi
}%
\providecommand \@ifx [1]{%
 \ifx #1\expandafter \@firstoftwo
 \else \expandafter \@secondoftwo
 \fi
}%
\providecommand \natexlab [1]{#1}%
\providecommand \enquote  [1]{``#1''}%
\providecommand \bibnamefont  [1]{#1}%
\providecommand \bibfnamefont [1]{#1}%
\providecommand \citenamefont [1]{#1}%
\providecommand \href@noop [0]{\@secondoftwo}%
\providecommand \href [0]{\begingroup \@sanitize@url \@href}%
\providecommand \@href[1]{\@@startlink{#1}\@@href}%
\providecommand \@@href[1]{\endgroup#1\@@endlink}%
\providecommand \@sanitize@url [0]{\catcode `\\12\catcode `\$12\catcode
  `\&12\catcode `\#12\catcode `\^12\catcode `\_12\catcode `\%12\relax}%
\providecommand \@@startlink[1]{}%
\providecommand \@@endlink[0]{}%
\providecommand \url  [0]{\begingroup\@sanitize@url \@url }%
\providecommand \@url [1]{\endgroup\@href {#1}{\urlprefix }}%
\providecommand \urlprefix  [0]{URL }%
\providecommand \Eprint [0]{\href }%
\providecommand \doibase [0]{https://doi.org/}%
\providecommand \selectlanguage [0]{\@gobble}%
\providecommand \bibinfo  [0]{\@secondoftwo}%
\providecommand \bibfield  [0]{\@secondoftwo}%
\providecommand \translation [1]{[#1]}%
\providecommand \BibitemOpen [0]{}%
\providecommand \bibitemStop [0]{}%
\providecommand \bibitemNoStop [0]{.\EOS\space}%
\providecommand \EOS [0]{\spacefactor3000\relax}%
\providecommand \BibitemShut  [1]{\csname bibitem#1\endcsname}%
\let\auto@bib@innerbib\@empty
\bibitem [{\citenamefont {Loss}\ and\ \citenamefont
  {DiVincenzo}(1998)}]{loss1998}%
  \BibitemOpen
  \bibfield  {author} {\bibinfo {author} {\bibfnamefont {D.}~\bibnamefont
  {Loss}}\ and\ \bibinfo {author} {\bibfnamefont {D.~P.}\ \bibnamefont
  {DiVincenzo}},\ }\bibfield  {title} {\bibinfo {title} {Quantum computation
  with quantum dots},\ }\href {https://doi.org/10.1103/PhysRevA.57.120}
  {\bibfield  {journal} {\bibinfo  {journal} {Phys. Rev. A}\ }\textbf {\bibinfo
  {volume} {57}},\ \bibinfo {pages} {120} (\bibinfo {year} {1998})}\BibitemShut
  {NoStop}%
\bibitem [{\citenamefont {van~der Wiel}\ \emph {et~al.}(2002)\citenamefont
  {van~der Wiel}, \citenamefont {De~Franceschi}, \citenamefont {Elzerman},
  \citenamefont {Fujisawa}, \citenamefont {Tarucha},\ and\ \citenamefont
  {Kouwenhoven}}]{vanderwiel2002}%
  \BibitemOpen
  \bibfield  {author} {\bibinfo {author} {\bibfnamefont {W.~G.}\ \bibnamefont
  {van~der Wiel}}, \bibinfo {author} {\bibfnamefont {S.}~\bibnamefont
  {De~Franceschi}}, \bibinfo {author} {\bibfnamefont {J.~M.}\ \bibnamefont
  {Elzerman}}, \bibinfo {author} {\bibfnamefont {T.}~\bibnamefont {Fujisawa}},
  \bibinfo {author} {\bibfnamefont {S.}~\bibnamefont {Tarucha}},\ and\ \bibinfo
  {author} {\bibfnamefont {L.~P.}\ \bibnamefont {Kouwenhoven}},\ }\bibfield
  {title} {\bibinfo {title} {Electron transport through double quantum dots},\
  }\href {https://doi.org/10.1103/RevModPhys.75.1} {\bibfield  {journal}
  {\bibinfo  {journal} {Rev. Mod. Phys.}\ }\textbf {\bibinfo {volume} {75}},\
  \bibinfo {pages} {1} (\bibinfo {year} {2002})}\BibitemShut {NoStop}%
\bibitem [{\citenamefont {Hanson}\ \emph {et~al.}(2007)\citenamefont {Hanson},
  \citenamefont {Kouwenhoven}, \citenamefont {Petta}, \citenamefont {Tarucha},\
  and\ \citenamefont {Vandersypen}}]{hanson2007}%
  \BibitemOpen
  \bibfield  {author} {\bibinfo {author} {\bibfnamefont {R.}~\bibnamefont
  {Hanson}}, \bibinfo {author} {\bibfnamefont {L.~P.}\ \bibnamefont
  {Kouwenhoven}}, \bibinfo {author} {\bibfnamefont {J.~R.}\ \bibnamefont
  {Petta}}, \bibinfo {author} {\bibfnamefont {S.}~\bibnamefont {Tarucha}},\
  and\ \bibinfo {author} {\bibfnamefont {L.~M.~K.}\ \bibnamefont
  {Vandersypen}},\ }\bibfield  {title} {\bibinfo {title} {Spins in few-electron
  quantum dots},\ }\href {https://doi.org/10.1103/RevModPhys.79.1217}
  {\bibfield  {journal} {\bibinfo  {journal} {Rev. Mod. Phys.}\ }\textbf
  {\bibinfo {volume} {79}},\ \bibinfo {pages} {1217} (\bibinfo {year}
  {2007})}\BibitemShut {NoStop}%
\bibitem [{\citenamefont {Ladd}\ \emph {et~al.}(2010)\citenamefont {Ladd},
  \citenamefont {Jelezko}, \citenamefont {Laflamme}, \citenamefont {Nakamura},
  \citenamefont {Monroe},\ and\ \citenamefont {O’Brien}}]{ladd2010}%
  \BibitemOpen
  \bibfield  {author} {\bibinfo {author} {\bibfnamefont {T.~D.}\ \bibnamefont
  {Ladd}}, \bibinfo {author} {\bibfnamefont {F.}~\bibnamefont {Jelezko}},
  \bibinfo {author} {\bibfnamefont {R.}~\bibnamefont {Laflamme}}, \bibinfo
  {author} {\bibfnamefont {Y.}~\bibnamefont {Nakamura}}, \bibinfo {author}
  {\bibfnamefont {C.}~\bibnamefont {Monroe}},\ and\ \bibinfo {author}
  {\bibfnamefont {J.~L.}\ \bibnamefont {O’Brien}},\ }\bibfield  {title}
  {\bibinfo {title} {Quantum computers},\ }\href
  {https://doi.org/10.1038/nature08812} {\bibfield  {journal} {\bibinfo
  {journal} {Nature}\ }\textbf {\bibinfo {volume} {464}},\ \bibinfo {pages}
  {45} (\bibinfo {year} {2010})}\BibitemShut {NoStop}%
\bibitem [{\citenamefont {Zwanenburg}\ \emph {et~al.}(2013)\citenamefont
  {Zwanenburg}, \citenamefont {Dzurak}, \citenamefont {Morello}, \citenamefont
  {Simmons}, \citenamefont {Hollenberg}, \citenamefont {Klimeck}, \citenamefont
  {Rogge}, \citenamefont {Coppersmith},\ and\ \citenamefont
  {Eriksson}}]{zwanenburg2013}%
  \BibitemOpen
  \bibfield  {author} {\bibinfo {author} {\bibfnamefont {F.~A.}\ \bibnamefont
  {Zwanenburg}}, \bibinfo {author} {\bibfnamefont {A.~S.}\ \bibnamefont
  {Dzurak}}, \bibinfo {author} {\bibfnamefont {A.}~\bibnamefont {Morello}},
  \bibinfo {author} {\bibfnamefont {M.~Y.}\ \bibnamefont {Simmons}}, \bibinfo
  {author} {\bibfnamefont {L.~C.~L.}\ \bibnamefont {Hollenberg}}, \bibinfo
  {author} {\bibfnamefont {G.}~\bibnamefont {Klimeck}}, \bibinfo {author}
  {\bibfnamefont {S.}~\bibnamefont {Rogge}}, \bibinfo {author} {\bibfnamefont
  {S.~N.}\ \bibnamefont {Coppersmith}},\ and\ \bibinfo {author} {\bibfnamefont
  {M.~A.}\ \bibnamefont {Eriksson}},\ }\bibfield  {title} {\bibinfo {title}
  {Silicon quantum electronics},\ }\href
  {https://doi.org/10.1103/RevModPhys.85.961} {\bibfield  {journal} {\bibinfo
  {journal} {Rev. Mod. Phys.}\ }\textbf {\bibinfo {volume} {85}},\ \bibinfo
  {pages} {961} (\bibinfo {year} {2013})}\BibitemShut {NoStop}%
\bibitem [{\citenamefont {Ladd}\ and\ \citenamefont
  {Carroll}(2018)}]{ladd2018}%
  \BibitemOpen
  \bibfield  {author} {\bibinfo {author} {\bibfnamefont {T.~D.}\ \bibnamefont
  {Ladd}}\ and\ \bibinfo {author} {\bibfnamefont {M.~S.}\ \bibnamefont
  {Carroll}},\ }\bibfield  {title} {\bibinfo {title} {Silicon qubits},\ }in\
  \href {https://doi.org/10.1016/B978-0-12-803581-8.09736-8} {\emph {\bibinfo
  {booktitle} {Encyclopedia of {Modern} {Optics} ({Second} {Edition})}}},\
  \bibinfo {editor} {edited by\ \bibinfo {editor} {\bibfnamefont {B.~D.}\
  \bibnamefont {Guenther}}\ and\ \bibinfo {editor} {\bibfnamefont {D.~G.}\
  \bibnamefont {Steel}}}\ (\bibinfo  {publisher} {Elsevier},\ \bibinfo
  {address} {Oxford},\ \bibinfo {year} {2018})\ pp.\ \bibinfo {pages}
  {467--477}\BibitemShut {NoStop}%
\bibitem [{\citenamefont {Eriksson}\ \emph {et~al.}(2004)\citenamefont
  {Eriksson}, \citenamefont {Friesen}, \citenamefont {Coppersmith},
  \citenamefont {Joynt}, \citenamefont {Klein}, \citenamefont {Slinker},
  \citenamefont {Tahan}, \citenamefont {Mooney}, \citenamefont {Chu},\ and\
  \citenamefont {Koester}}]{eriksson2004}%
  \BibitemOpen
  \bibfield  {author} {\bibinfo {author} {\bibfnamefont {M.~A.}\ \bibnamefont
  {Eriksson}}, \bibinfo {author} {\bibfnamefont {M.}~\bibnamefont {Friesen}},
  \bibinfo {author} {\bibfnamefont {S.~N.}\ \bibnamefont {Coppersmith}},
  \bibinfo {author} {\bibfnamefont {R.}~\bibnamefont {Joynt}}, \bibinfo
  {author} {\bibfnamefont {L.~J.}\ \bibnamefont {Klein}}, \bibinfo {author}
  {\bibfnamefont {K.}~\bibnamefont {Slinker}}, \bibinfo {author} {\bibfnamefont
  {C.}~\bibnamefont {Tahan}}, \bibinfo {author} {\bibfnamefont {P.~M.}\
  \bibnamefont {Mooney}}, \bibinfo {author} {\bibfnamefont {J.~O.}\
  \bibnamefont {Chu}},\ and\ \bibinfo {author} {\bibfnamefont {S.~J.}\
  \bibnamefont {Koester}},\ }\bibfield  {title} {\bibinfo {title} {Spin-based
  quantum dot quantum computing in silicon},\ }\href
  {https://doi.org/10.1007/s11128-004-2224-z} {\bibfield  {journal} {\bibinfo
  {journal} {Quant. Inf. Proc.}\ }\textbf {\bibinfo {volume} {3}},\ \bibinfo
  {pages} {133} (\bibinfo {year} {2004})}\BibitemShut {NoStop}%
\bibitem [{\citenamefont {Maune}\ \emph {et~al.}(2012)\citenamefont {Maune},
  \citenamefont {Borselli}, \citenamefont {Huang}, \citenamefont {Ladd},
  \citenamefont {Deelman}, \citenamefont {Holabird}, \citenamefont {Kiselev},
  \citenamefont {Alvarado-Rodriguez}, \citenamefont {Ross}, \citenamefont
  {Schmitz}, \citenamefont {Sokolich}, \citenamefont {Watson}, \citenamefont
  {Gyure},\ and\ \citenamefont {Hunter}}]{maune2012}%
  \BibitemOpen
  \bibfield  {author} {\bibinfo {author} {\bibfnamefont {B.~M.}\ \bibnamefont
  {Maune}}, \bibinfo {author} {\bibfnamefont {M.~G.}\ \bibnamefont {Borselli}},
  \bibinfo {author} {\bibfnamefont {B.}~\bibnamefont {Huang}}, \bibinfo
  {author} {\bibfnamefont {T.~D.}\ \bibnamefont {Ladd}}, \bibinfo {author}
  {\bibfnamefont {P.~W.}\ \bibnamefont {Deelman}}, \bibinfo {author}
  {\bibfnamefont {K.~S.}\ \bibnamefont {Holabird}}, \bibinfo {author}
  {\bibfnamefont {A.~A.}\ \bibnamefont {Kiselev}}, \bibinfo {author}
  {\bibfnamefont {I.}~\bibnamefont {Alvarado-Rodriguez}}, \bibinfo {author}
  {\bibfnamefont {R.~S.}\ \bibnamefont {Ross}}, \bibinfo {author}
  {\bibfnamefont {A.~E.}\ \bibnamefont {Schmitz}}, \bibinfo {author}
  {\bibfnamefont {M.}~\bibnamefont {Sokolich}}, \bibinfo {author}
  {\bibfnamefont {C.~A.}\ \bibnamefont {Watson}}, \bibinfo {author}
  {\bibfnamefont {M.~F.}\ \bibnamefont {Gyure}},\ and\ \bibinfo {author}
  {\bibfnamefont {A.~T.}\ \bibnamefont {Hunter}},\ }\bibfield  {title}
  {\bibinfo {title} {Coherent singlet-triplet oscillations in a silicon-based
  double quantum dot},\ }\href {http://dx.doi.org/10.1038/nature10707}
  {\bibfield  {journal} {\bibinfo  {journal} {Nature}\ }\textbf {\bibinfo
  {volume} {481}},\ \bibinfo {pages} {344} (\bibinfo {year}
  {2012})}\BibitemShut {NoStop}%
\bibitem [{\citenamefont {Reed}\ \emph {et~al.}(2016)\citenamefont {Reed},
  \citenamefont {Maune}, \citenamefont {Andrews}, \citenamefont {Borselli},
  \citenamefont {Eng}, \citenamefont {Jura}, \citenamefont {Kiselev},
  \citenamefont {Ladd}, \citenamefont {Merkel}, \citenamefont {Milosavljevic},
  \citenamefont {Pritchett}, \citenamefont {Rakher}, \citenamefont {Ross},
  \citenamefont {Schmitz}, \citenamefont {Smith}, \citenamefont {Wright},
  \citenamefont {Gyure},\ and\ \citenamefont {Hunter}}]{reed2016}%
  \BibitemOpen
  \bibfield  {author} {\bibinfo {author} {\bibfnamefont {M.~D.}\ \bibnamefont
  {Reed}}, \bibinfo {author} {\bibfnamefont {B.~M.}\ \bibnamefont {Maune}},
  \bibinfo {author} {\bibfnamefont {R.~W.}\ \bibnamefont {Andrews}}, \bibinfo
  {author} {\bibfnamefont {M.~G.}\ \bibnamefont {Borselli}}, \bibinfo {author}
  {\bibfnamefont {K.}~\bibnamefont {Eng}}, \bibinfo {author} {\bibfnamefont
  {M.~P.}\ \bibnamefont {Jura}}, \bibinfo {author} {\bibfnamefont {A.~A.}\
  \bibnamefont {Kiselev}}, \bibinfo {author} {\bibfnamefont {T.~D.}\
  \bibnamefont {Ladd}}, \bibinfo {author} {\bibfnamefont {S.~T.}\ \bibnamefont
  {Merkel}}, \bibinfo {author} {\bibfnamefont {I.}~\bibnamefont
  {Milosavljevic}}, \bibinfo {author} {\bibfnamefont {E.~J.}\ \bibnamefont
  {Pritchett}}, \bibinfo {author} {\bibfnamefont {M.~T.}\ \bibnamefont
  {Rakher}}, \bibinfo {author} {\bibfnamefont {R.~S.}\ \bibnamefont {Ross}},
  \bibinfo {author} {\bibfnamefont {A.~E.}\ \bibnamefont {Schmitz}}, \bibinfo
  {author} {\bibfnamefont {A.}~\bibnamefont {Smith}}, \bibinfo {author}
  {\bibfnamefont {J.~A.}\ \bibnamefont {Wright}}, \bibinfo {author}
  {\bibfnamefont {M.~F.}\ \bibnamefont {Gyure}},\ and\ \bibinfo {author}
  {\bibfnamefont {A.~T.}\ \bibnamefont {Hunter}},\ }\bibfield  {title}
  {\bibinfo {title} {Reduced sensitivity to charge noise in semiconductor spin
  qubits via symmetric operation},\ }\href
  {https://doi.org/10.1103/PhysRevLett.116.110402} {\bibfield  {journal}
  {\bibinfo  {journal} {Phys. Rev. Lett.}\ }\textbf {\bibinfo {volume} {116}},\
  \bibinfo {pages} {110402} (\bibinfo {year} {2016})}\BibitemShut {NoStop}%
\bibitem [{\citenamefont {Kawakami}\ \emph {et~al.}(2016)\citenamefont
  {Kawakami}, \citenamefont {Jullien}, \citenamefont {Scarlino}, \citenamefont
  {Ward}, \citenamefont {Savage}, \citenamefont {Lagally}, \citenamefont
  {Dobrovitski}, \citenamefont {Friesen}, \citenamefont {Coppersmith},
  \citenamefont {Eriksson},\ and\ \citenamefont {Vandersypen}}]{kawakami2016}%
  \BibitemOpen
  \bibfield  {author} {\bibinfo {author} {\bibfnamefont {E.}~\bibnamefont
  {Kawakami}}, \bibinfo {author} {\bibfnamefont {T.}~\bibnamefont {Jullien}},
  \bibinfo {author} {\bibfnamefont {P.}~\bibnamefont {Scarlino}}, \bibinfo
  {author} {\bibfnamefont {D.~R.}\ \bibnamefont {Ward}}, \bibinfo {author}
  {\bibfnamefont {D.~E.}\ \bibnamefont {Savage}}, \bibinfo {author}
  {\bibfnamefont {M.~G.}\ \bibnamefont {Lagally}}, \bibinfo {author}
  {\bibfnamefont {V.~V.}\ \bibnamefont {Dobrovitski}}, \bibinfo {author}
  {\bibfnamefont {M.}~\bibnamefont {Friesen}}, \bibinfo {author} {\bibfnamefont
  {S.~N.}\ \bibnamefont {Coppersmith}}, \bibinfo {author} {\bibfnamefont
  {M.~A.}\ \bibnamefont {Eriksson}},\ and\ \bibinfo {author} {\bibfnamefont
  {L.~M.~K.}\ \bibnamefont {Vandersypen}},\ }\bibfield  {title} {\bibinfo
  {title} {Gate fidelity and coherence of an electron spin in an {Si/SiGe}
  quantum dot with micromagnet},\ }\href
  {https://doi.org/10.1073/pnas.1603251113} {\bibfield  {journal} {\bibinfo
  {journal} {Proc. Nat. Acad. Sci. USA}\ }\textbf {\bibinfo {volume} {113}},\
  \bibinfo {pages} {11738} (\bibinfo {year} {2016})}\BibitemShut {NoStop}%
\bibitem [{\citenamefont {Connors}\ \emph {et~al.}(2019)\citenamefont
  {Connors}, \citenamefont {Nelson}, \citenamefont {Qiao}, \citenamefont
  {Edge},\ and\ \citenamefont {Nichol}}]{connors2019}%
  \BibitemOpen
  \bibfield  {author} {\bibinfo {author} {\bibfnamefont {E.~J.}\ \bibnamefont
  {Connors}}, \bibinfo {author} {\bibfnamefont {J.}~\bibnamefont {Nelson}},
  \bibinfo {author} {\bibfnamefont {H.}~\bibnamefont {Qiao}}, \bibinfo {author}
  {\bibfnamefont {L.~F.}\ \bibnamefont {Edge}},\ and\ \bibinfo {author}
  {\bibfnamefont {J.~M.}\ \bibnamefont {Nichol}},\ }\bibfield  {title}
  {\bibinfo {title} {Low-frequency charge noise in {Si/SiGe} quantum dots},\
  }\href {https://doi.org/10.1103/PhysRevB.100.165305} {\bibfield  {journal}
  {\bibinfo  {journal} {Phys. Rev. B}\ }\textbf {\bibinfo {volume} {100}},\
  \bibinfo {pages} {165305} (\bibinfo {year} {2019})}\BibitemShut {NoStop}%
\bibitem [{\citenamefont {Tyryshkin}\ \emph {et~al.}(2012)\citenamefont
  {Tyryshkin}, \citenamefont {Tojo}, \citenamefont {Morton}, \citenamefont
  {Riemann}, \citenamefont {Abrosimov}, \citenamefont {Becker}, \citenamefont
  {Pohl}, \citenamefont {Schenkel}, \citenamefont {Thewalt}, \citenamefont
  {Itoh},\ and\ \citenamefont {Lyon}}]{tyryshkin2012}%
  \BibitemOpen
  \bibfield  {author} {\bibinfo {author} {\bibfnamefont {A.~M.}\ \bibnamefont
  {Tyryshkin}}, \bibinfo {author} {\bibfnamefont {S.}~\bibnamefont {Tojo}},
  \bibinfo {author} {\bibfnamefont {J.~J.~L.}\ \bibnamefont {Morton}}, \bibinfo
  {author} {\bibfnamefont {H.}~\bibnamefont {Riemann}}, \bibinfo {author}
  {\bibfnamefont {N.~V.}\ \bibnamefont {Abrosimov}}, \bibinfo {author}
  {\bibfnamefont {P.}~\bibnamefont {Becker}}, \bibinfo {author} {\bibfnamefont
  {H.-J.}\ \bibnamefont {Pohl}}, \bibinfo {author} {\bibfnamefont
  {T.}~\bibnamefont {Schenkel}}, \bibinfo {author} {\bibfnamefont {M.~L.~W.}\
  \bibnamefont {Thewalt}}, \bibinfo {author} {\bibfnamefont {K.~M.}\
  \bibnamefont {Itoh}},\ and\ \bibinfo {author} {\bibfnamefont {S.~A.}\
  \bibnamefont {Lyon}},\ }\bibfield  {title} {\bibinfo {title} {Electron spin
  coherence exceeding seconds in high-purity silicon},\ }\href
  {https://doi.org/10.1038/nmat3182} {\bibfield  {journal} {\bibinfo  {journal}
  {Nat. Mater.}\ }\textbf {\bibinfo {volume} {11}},\ \bibinfo {pages} {143}
  (\bibinfo {year} {2012})}\BibitemShut {NoStop}%
\bibitem [{\citenamefont {Eng}\ \emph {et~al.}(2015)\citenamefont {Eng},
  \citenamefont {Ladd}, \citenamefont {Smith}, \citenamefont {Borselli},
  \citenamefont {Kiselev}, \citenamefont {Fong}, \citenamefont {Holabird},
  \citenamefont {Hazard}, \citenamefont {Huang}, \citenamefont {Deelman},
  \citenamefont {Milosavljevic}, \citenamefont {Schmitz}, \citenamefont {Ross},
  \citenamefont {Gyure},\ and\ \citenamefont {Hunter}}]{eng2015}%
  \BibitemOpen
  \bibfield  {author} {\bibinfo {author} {\bibfnamefont {K.}~\bibnamefont
  {Eng}}, \bibinfo {author} {\bibfnamefont {T.~D.}\ \bibnamefont {Ladd}},
  \bibinfo {author} {\bibfnamefont {A.}~\bibnamefont {Smith}}, \bibinfo
  {author} {\bibfnamefont {M.~G.}\ \bibnamefont {Borselli}}, \bibinfo {author}
  {\bibfnamefont {A.~A.}\ \bibnamefont {Kiselev}}, \bibinfo {author}
  {\bibfnamefont {B.~H.}\ \bibnamefont {Fong}}, \bibinfo {author}
  {\bibfnamefont {K.~S.}\ \bibnamefont {Holabird}}, \bibinfo {author}
  {\bibfnamefont {T.~M.}\ \bibnamefont {Hazard}}, \bibinfo {author}
  {\bibfnamefont {B.}~\bibnamefont {Huang}}, \bibinfo {author} {\bibfnamefont
  {P.~W.}\ \bibnamefont {Deelman}}, \bibinfo {author} {\bibfnamefont
  {I.}~\bibnamefont {Milosavljevic}}, \bibinfo {author} {\bibfnamefont {A.~E.}\
  \bibnamefont {Schmitz}}, \bibinfo {author} {\bibfnamefont {R.~S.}\
  \bibnamefont {Ross}}, \bibinfo {author} {\bibfnamefont {M.~F.}\ \bibnamefont
  {Gyure}},\ and\ \bibinfo {author} {\bibfnamefont {A.~T.}\ \bibnamefont
  {Hunter}},\ }\bibfield  {title} {\bibinfo {title} {Isotopically enhanced
  triple-quantum-dot qubit},\ }\href {https://doi.org/10.1126/sciadv.1500214}
  {\bibfield  {journal} {\bibinfo  {journal} {Sci. Adv.}\ }\textbf {\bibinfo
  {volume} {1}},\ \bibinfo {pages} {e1500214} (\bibinfo {year}
  {2015})}\BibitemShut {NoStop}%
\bibitem [{\citenamefont {Struck}\ \emph {et~al.}(2020)\citenamefont {Struck},
  \citenamefont {Hollmann}, \citenamefont {Schauer}, \citenamefont {Fedorets},
  \citenamefont {Schmidbauer}, \citenamefont {Sawano}, \citenamefont {Riemann},
  \citenamefont {Abrosimov}, \citenamefont {Cywi{\'n}ski}, \citenamefont
  {Bougeard} \emph {et~al.}}]{struck2020}%
  \BibitemOpen
  \bibfield  {author} {\bibinfo {author} {\bibfnamefont {T.}~\bibnamefont
  {Struck}}, \bibinfo {author} {\bibfnamefont {A.}~\bibnamefont {Hollmann}},
  \bibinfo {author} {\bibfnamefont {F.}~\bibnamefont {Schauer}}, \bibinfo
  {author} {\bibfnamefont {O.}~\bibnamefont {Fedorets}}, \bibinfo {author}
  {\bibfnamefont {A.}~\bibnamefont {Schmidbauer}}, \bibinfo {author}
  {\bibfnamefont {K.}~\bibnamefont {Sawano}}, \bibinfo {author} {\bibfnamefont
  {H.}~\bibnamefont {Riemann}}, \bibinfo {author} {\bibfnamefont {N.~V.}\
  \bibnamefont {Abrosimov}}, \bibinfo {author} {\bibfnamefont
  {{\L}.}~\bibnamefont {Cywi{\'n}ski}}, \bibinfo {author} {\bibfnamefont
  {D.}~\bibnamefont {Bougeard}}, \emph {et~al.},\ }\bibfield  {title} {\bibinfo
  {title} {Low-frequency spin qubit energy splitting noise in highly purified
  {$^{28}$Si/SiGe}},\ }\href {https://doi.org/10.1038/s41534-020-0276-2}
  {\bibfield  {journal} {\bibinfo  {journal} {npj Quantum Inf.}\ }\textbf
  {\bibinfo {volume} {6}},\ \bibinfo {pages} {1} (\bibinfo {year}
  {2020})}\BibitemShut {NoStop}%
\bibitem [{\citenamefont {Lidar}\ \emph {et~al.}(1998)\citenamefont {Lidar},
  \citenamefont {Chuang},\ and\ \citenamefont {Whaley}}]{lidar1998}%
  \BibitemOpen
  \bibfield  {author} {\bibinfo {author} {\bibfnamefont {D.~A.}\ \bibnamefont
  {Lidar}}, \bibinfo {author} {\bibfnamefont {I.~L.}\ \bibnamefont {Chuang}},\
  and\ \bibinfo {author} {\bibfnamefont {K.~B.}\ \bibnamefont {Whaley}},\
  }\bibfield  {title} {\bibinfo {title} {Decoherence-free subspaces for quantum
  computation},\ }\href {https://doi.org/10.1103/PhysRevLett.81.2594}
  {\bibfield  {journal} {\bibinfo  {journal} {Phys. Rev. Lett.}\ }\textbf
  {\bibinfo {volume} {81}},\ \bibinfo {pages} {2594} (\bibinfo {year}
  {1998})}\BibitemShut {NoStop}%
\bibitem [{\citenamefont {DiVincenzo}\ \emph {et~al.}(2000)\citenamefont
  {DiVincenzo}, \citenamefont {Bacon}, \citenamefont {Kempe}, \citenamefont
  {Burkard},\ and\ \citenamefont {Whaley}}]{divincenzo2000}%
  \BibitemOpen
  \bibfield  {author} {\bibinfo {author} {\bibfnamefont {D.~P.}\ \bibnamefont
  {DiVincenzo}}, \bibinfo {author} {\bibfnamefont {D.}~\bibnamefont {Bacon}},
  \bibinfo {author} {\bibfnamefont {J.}~\bibnamefont {Kempe}}, \bibinfo
  {author} {\bibfnamefont {G.}~\bibnamefont {Burkard}},\ and\ \bibinfo {author}
  {\bibfnamefont {K.~B.}\ \bibnamefont {Whaley}},\ }\bibfield  {title}
  {\bibinfo {title} {Universal quantum computation with the exchange
  interaction},\ }\href {http://dx.doi.org/10.1038/35042541} {\bibfield
  {journal} {\bibinfo  {journal} {Nature}\ }\textbf {\bibinfo {volume} {408}},\
  \bibinfo {pages} {339} (\bibinfo {year} {2000})}\BibitemShut {NoStop}%
\bibitem [{\citenamefont {Kempe}\ \emph {et~al.}(2001)\citenamefont {Kempe},
  \citenamefont {Bacon}, \citenamefont {Lidar},\ and\ \citenamefont
  {Whaley}}]{kempe2001}%
  \BibitemOpen
  \bibfield  {author} {\bibinfo {author} {\bibfnamefont {J.}~\bibnamefont
  {Kempe}}, \bibinfo {author} {\bibfnamefont {D.}~\bibnamefont {Bacon}},
  \bibinfo {author} {\bibfnamefont {D.~A.}\ \bibnamefont {Lidar}},\ and\
  \bibinfo {author} {\bibfnamefont {K.~B.}\ \bibnamefont {Whaley}},\ }\bibfield
   {title} {\bibinfo {title} {Theory of decoherence-free fault-tolerant
  universal quantum computation},\ }\href
  {https://doi.org/10.1103/PhysRevA.63.042307} {\bibfield  {journal} {\bibinfo
  {journal} {Phys. Rev. A}\ }\textbf {\bibinfo {volume} {63}},\ \bibinfo
  {pages} {042307} (\bibinfo {year} {2001})}\BibitemShut {NoStop}%
\bibitem [{\citenamefont {Fong}\ and\ \citenamefont
  {Wandzura}(2011)}]{fong2011}%
  \BibitemOpen
  \bibfield  {author} {\bibinfo {author} {\bibfnamefont {B.~H.}\ \bibnamefont
  {Fong}}\ and\ \bibinfo {author} {\bibfnamefont {S.~M.}\ \bibnamefont
  {Wandzura}},\ }\bibfield  {title} {\bibinfo {title} {Universal quantum
  computation and leakage reduction in the 3-qubit decoherence free
  subsystem},\ }\href {http://dl.acm.org/citation.cfm?id=2230956.2230965}
  {\bibfield  {journal} {\bibinfo  {journal} {Quantum Info. Comput.}\ }\textbf
  {\bibinfo {volume} {11}},\ \bibinfo {pages} {1003} (\bibinfo {year}
  {2011})}\BibitemShut {NoStop}%
\bibitem [{\citenamefont {Andrews}\ \emph {et~al.}(2019)\citenamefont
  {Andrews}, \citenamefont {Jones}, \citenamefont {Reed}, \citenamefont
  {Jones}, \citenamefont {Ha}, \citenamefont {Jura}, \citenamefont {Kerckhoff},
  \citenamefont {Levendorf}, \citenamefont {Meenehan}, \citenamefont {Merkel},
  \citenamefont {Smith}, \citenamefont {Sun}, \citenamefont {Weinstein},
  \citenamefont {Rakher}, \citenamefont {Ladd},\ and\ \citenamefont
  {Borselli}}]{andrews2019}%
  \BibitemOpen
  \bibfield  {author} {\bibinfo {author} {\bibfnamefont {R.~W.}\ \bibnamefont
  {Andrews}}, \bibinfo {author} {\bibfnamefont {C.}~\bibnamefont {Jones}},
  \bibinfo {author} {\bibfnamefont {M.~D.}\ \bibnamefont {Reed}}, \bibinfo
  {author} {\bibfnamefont {A.~M.}\ \bibnamefont {Jones}}, \bibinfo {author}
  {\bibfnamefont {S.~D.}\ \bibnamefont {Ha}}, \bibinfo {author} {\bibfnamefont
  {M.~P.}\ \bibnamefont {Jura}}, \bibinfo {author} {\bibfnamefont
  {J.}~\bibnamefont {Kerckhoff}}, \bibinfo {author} {\bibfnamefont
  {M.}~\bibnamefont {Levendorf}}, \bibinfo {author} {\bibfnamefont
  {S.}~\bibnamefont {Meenehan}}, \bibinfo {author} {\bibfnamefont {S.~T.}\
  \bibnamefont {Merkel}}, \bibinfo {author} {\bibfnamefont {A.}~\bibnamefont
  {Smith}}, \bibinfo {author} {\bibfnamefont {B.}~\bibnamefont {Sun}}, \bibinfo
  {author} {\bibfnamefont {A.~J.}\ \bibnamefont {Weinstein}}, \bibinfo {author}
  {\bibfnamefont {M.~T.}\ \bibnamefont {Rakher}}, \bibinfo {author}
  {\bibfnamefont {T.~D.}\ \bibnamefont {Ladd}},\ and\ \bibinfo {author}
  {\bibfnamefont {M.~G.}\ \bibnamefont {Borselli}},\ }\bibfield  {title}
  {\bibinfo {title} {Quantifying error and leakage in an encoded {Si}/{SiGe}
  triple-dot qubit},\ }\href {https://doi.org/10.1038/s41565-019-0500-4}
  {\bibfield  {journal} {\bibinfo  {journal} {Nat. Nanotechnol.}\ }\textbf
  {\bibinfo {volume} {14}},\ \bibinfo {pages} {747} (\bibinfo {year}
  {2019})}\BibitemShut {NoStop}%
\bibitem [{\citenamefont {Elzerman}\ \emph {et~al.}(2004)\citenamefont
  {Elzerman}, \citenamefont {Hanson}, \citenamefont {Willems~van Beveren},
  \citenamefont {Witkamp}, \citenamefont {Vandersypen},\ and\ \citenamefont
  {Kouwenhoven}}]{elzerman2004b}%
  \BibitemOpen
  \bibfield  {author} {\bibinfo {author} {\bibfnamefont {J.~M.}\ \bibnamefont
  {Elzerman}}, \bibinfo {author} {\bibfnamefont {R.}~\bibnamefont {Hanson}},
  \bibinfo {author} {\bibfnamefont {L.~H.}\ \bibnamefont {Willems~van
  Beveren}}, \bibinfo {author} {\bibfnamefont {B.}~\bibnamefont {Witkamp}},
  \bibinfo {author} {\bibfnamefont {L.~M.~K.}\ \bibnamefont {Vandersypen}},\
  and\ \bibinfo {author} {\bibfnamefont {L.~P.}\ \bibnamefont {Kouwenhoven}},\
  }\bibfield  {title} {\bibinfo {title} {Single-shot read-out of an individual
  electron spin in a quantum dot},\ }\href
  {https://doi.org/10.1038/nature02693} {\bibfield  {journal} {\bibinfo
  {journal} {Nature}\ }\textbf {\bibinfo {volume} {430}},\ \bibinfo {pages}
  {431} (\bibinfo {year} {2004})}\BibitemShut {NoStop}%
\bibitem [{\citenamefont {Ono}\ \emph {et~al.}(2002)\citenamefont {Ono},
  \citenamefont {Austing}, \citenamefont {Tokura},\ and\ \citenamefont
  {Tarucha}}]{ono2002}%
  \BibitemOpen
  \bibfield  {author} {\bibinfo {author} {\bibfnamefont {K.}~\bibnamefont
  {Ono}}, \bibinfo {author} {\bibfnamefont {D.}~\bibnamefont {Austing}},
  \bibinfo {author} {\bibfnamefont {Y.}~\bibnamefont {Tokura}},\ and\ \bibinfo
  {author} {\bibfnamefont {S.}~\bibnamefont {Tarucha}},\ }\bibfield  {title}
  {\bibinfo {title} {Current rectification by pauli exclusion in a weakly
  coupled double quantum dot system},\ }\href
  {https://doi.org/10.1126/science.1070958} {\bibfield  {journal} {\bibinfo
  {journal} {Science}\ }\textbf {\bibinfo {volume} {297}},\ \bibinfo {pages}
  {1313} (\bibinfo {year} {2002})}\BibitemShut {NoStop}%
\bibitem [{\citenamefont {Johnson}\ \emph {et~al.}(2005)\citenamefont
  {Johnson}, \citenamefont {Petta}, \citenamefont {Marcus}, \citenamefont
  {Hanson},\ and\ \citenamefont {Gossard}}]{johnson2005}%
  \BibitemOpen
  \bibfield  {author} {\bibinfo {author} {\bibfnamefont {A.~C.}\ \bibnamefont
  {Johnson}}, \bibinfo {author} {\bibfnamefont {J.~R.}\ \bibnamefont {Petta}},
  \bibinfo {author} {\bibfnamefont {C.~M.}\ \bibnamefont {Marcus}}, \bibinfo
  {author} {\bibfnamefont {M.~P.}\ \bibnamefont {Hanson}},\ and\ \bibinfo
  {author} {\bibfnamefont {A.~C.}\ \bibnamefont {Gossard}},\ }\bibfield
  {title} {\bibinfo {title} {Singlet-triplet spin blockade and charge sensing
  in a few-electron double quantum dot},\ }\href
  {https://doi.org/10.1103/PhysRevB.72.165308} {\bibfield  {journal} {\bibinfo
  {journal} {Phys. Rev. B}\ }\textbf {\bibinfo {volume} {72}},\ \bibinfo
  {pages} {165308} (\bibinfo {year} {2005})}\BibitemShut {NoStop}%
\bibitem [{\citenamefont {Field}\ \emph {et~al.}(1993)\citenamefont {Field},
  \citenamefont {Smith}, \citenamefont {Pepper}, \citenamefont {Ritchie},
  \citenamefont {Frost}, \citenamefont {Jones},\ and\ \citenamefont
  {Hasko}}]{field1993}%
  \BibitemOpen
  \bibfield  {author} {\bibinfo {author} {\bibfnamefont {M.}~\bibnamefont
  {Field}}, \bibinfo {author} {\bibfnamefont {C.~G.}\ \bibnamefont {Smith}},
  \bibinfo {author} {\bibfnamefont {M.}~\bibnamefont {Pepper}}, \bibinfo
  {author} {\bibfnamefont {D.~A.}\ \bibnamefont {Ritchie}}, \bibinfo {author}
  {\bibfnamefont {J.~E.~F.}\ \bibnamefont {Frost}}, \bibinfo {author}
  {\bibfnamefont {G.~A.~C.}\ \bibnamefont {Jones}},\ and\ \bibinfo {author}
  {\bibfnamefont {D.~G.}\ \bibnamefont {Hasko}},\ }\bibfield  {title} {\bibinfo
  {title} {Measurements of {Coulomb} blockade with a noninvasive voltage
  probe},\ }\href {https://doi.org/10.1103/PhysRevLett.70.1311} {\bibfield
  {journal} {\bibinfo  {journal} {Phys. Rev. Lett.}\ }\textbf {\bibinfo
  {volume} {70}},\ \bibinfo {pages} {1311} (\bibinfo {year}
  {1993})}\BibitemShut {NoStop}%
\bibitem [{\citenamefont {Kane}\ \emph {et~al.}(2000)\citenamefont {Kane},
  \citenamefont {McAlpine}, \citenamefont {Dzurak}, \citenamefont {Clark},
  \citenamefont {Milburn}, \citenamefont {Sun},\ and\ \citenamefont
  {Wiseman}}]{kane2004}%
  \BibitemOpen
  \bibfield  {author} {\bibinfo {author} {\bibfnamefont {B.~E.}\ \bibnamefont
  {Kane}}, \bibinfo {author} {\bibfnamefont {N.~S.}\ \bibnamefont {McAlpine}},
  \bibinfo {author} {\bibfnamefont {A.~S.}\ \bibnamefont {Dzurak}}, \bibinfo
  {author} {\bibfnamefont {R.~G.}\ \bibnamefont {Clark}}, \bibinfo {author}
  {\bibfnamefont {G.~J.}\ \bibnamefont {Milburn}}, \bibinfo {author}
  {\bibfnamefont {H.~B.}\ \bibnamefont {Sun}},\ and\ \bibinfo {author}
  {\bibfnamefont {H.}~\bibnamefont {Wiseman}},\ }\bibfield  {title} {\bibinfo
  {title} {Single-spin measurement using single-electron transistors to probe
  two-electron systems},\ }\href {https://doi.org/10.1103/PhysRevB.61.2961}
  {\bibfield  {journal} {\bibinfo  {journal} {Phys. Rev. B}\ }\textbf {\bibinfo
  {volume} {61}},\ \bibinfo {pages} {2961} (\bibinfo {year}
  {2000})}\BibitemShut {NoStop}%
\bibitem [{\citenamefont {Podd}\ \emph {et~al.}(2010)\citenamefont {Podd},
  \citenamefont {Angus}, \citenamefont {Williams},\ and\ \citenamefont
  {Ferguson}}]{podd2010}%
  \BibitemOpen
  \bibfield  {author} {\bibinfo {author} {\bibfnamefont {G.~J.}\ \bibnamefont
  {Podd}}, \bibinfo {author} {\bibfnamefont {S.~J.}\ \bibnamefont {Angus}},
  \bibinfo {author} {\bibfnamefont {D.~A.}\ \bibnamefont {Williams}},\ and\
  \bibinfo {author} {\bibfnamefont {A.~J.}\ \bibnamefont {Ferguson}},\
  }\bibfield  {title} {\bibinfo {title} {Charge sensing in intrinsic silicon
  quantum dots},\ }\href {https://doi.org/10.1063/1.3318463} {\bibfield
  {journal} {\bibinfo  {journal} {Appl. Phys. Lett.}\ }\textbf {\bibinfo
  {volume} {96}},\ \bibinfo {pages} {082104} (\bibinfo {year}
  {2010})}\BibitemShut {NoStop}%
\bibitem [{\citenamefont {Morello}\ \emph {et~al.}(2010)\citenamefont
  {Morello}, \citenamefont {Pla}, \citenamefont {Zwanenburg}, \citenamefont
  {Chan}, \citenamefont {Tan}, \citenamefont {Huebl}, \citenamefont
  {M{\"o}tt{\"o}nen}, \citenamefont {Nugroho}, \citenamefont {Yang},
  \citenamefont {van Donkelaar}, \citenamefont {Alves}, \citenamefont
  {Jamieson}, \citenamefont {Escott}, \citenamefont {Hollenberg}, \citenamefont
  {Clark},\ and\ \citenamefont {Dzurak}}]{morello2010}%
  \BibitemOpen
  \bibfield  {author} {\bibinfo {author} {\bibfnamefont {A.}~\bibnamefont
  {Morello}}, \bibinfo {author} {\bibfnamefont {J.~J.}\ \bibnamefont {Pla}},
  \bibinfo {author} {\bibfnamefont {F.~A.}\ \bibnamefont {Zwanenburg}},
  \bibinfo {author} {\bibfnamefont {K.~W.}\ \bibnamefont {Chan}}, \bibinfo
  {author} {\bibfnamefont {K.~Y.}\ \bibnamefont {Tan}}, \bibinfo {author}
  {\bibfnamefont {H.}~\bibnamefont {Huebl}}, \bibinfo {author} {\bibfnamefont
  {M.}~\bibnamefont {M{\"o}tt{\"o}nen}}, \bibinfo {author} {\bibfnamefont
  {C.~D.}\ \bibnamefont {Nugroho}}, \bibinfo {author} {\bibfnamefont
  {C.}~\bibnamefont {Yang}}, \bibinfo {author} {\bibfnamefont {J.~A.}\
  \bibnamefont {van Donkelaar}}, \bibinfo {author} {\bibfnamefont {A.~D.~C.}\
  \bibnamefont {Alves}}, \bibinfo {author} {\bibfnamefont {D.~N.}\ \bibnamefont
  {Jamieson}}, \bibinfo {author} {\bibfnamefont {C.~C.}\ \bibnamefont
  {Escott}}, \bibinfo {author} {\bibfnamefont {L.~C.~L.}\ \bibnamefont
  {Hollenberg}}, \bibinfo {author} {\bibfnamefont {R.~G.}\ \bibnamefont
  {Clark}},\ and\ \bibinfo {author} {\bibfnamefont {A.~S.}\ \bibnamefont
  {Dzurak}},\ }\bibfield  {title} {\bibinfo {title} {Single-shot readout of an
  electron spin in silicon},\ }\href {https://doi.org/10.1038/nature09392}
  {\bibfield  {journal} {\bibinfo  {journal} {Nature}\ }\textbf {\bibinfo
  {volume} {467}},\ \bibinfo {pages} {687} (\bibinfo {year}
  {2010})}\BibitemShut {NoStop}%
\bibitem [{\citenamefont {Dehollain}\ \emph {et~al.}(2014)\citenamefont
  {Dehollain}, \citenamefont {Muhonen}, \citenamefont {Tan}, \citenamefont
  {Saraiva}, \citenamefont {Jamieson}, \citenamefont {Dzurak},\ and\
  \citenamefont {Morello}}]{dehollain2014}%
  \BibitemOpen
  \bibfield  {author} {\bibinfo {author} {\bibfnamefont {J.~P.}\ \bibnamefont
  {Dehollain}}, \bibinfo {author} {\bibfnamefont {J.~T.}\ \bibnamefont
  {Muhonen}}, \bibinfo {author} {\bibfnamefont {K.~Y.}\ \bibnamefont {Tan}},
  \bibinfo {author} {\bibfnamefont {A.}~\bibnamefont {Saraiva}}, \bibinfo
  {author} {\bibfnamefont {D.~N.}\ \bibnamefont {Jamieson}}, \bibinfo {author}
  {\bibfnamefont {A.~S.}\ \bibnamefont {Dzurak}},\ and\ \bibinfo {author}
  {\bibfnamefont {A.}~\bibnamefont {Morello}},\ }\bibfield  {title} {\bibinfo
  {title} {Single-shot readout and relaxation of singlet and triplet states in
  exchange-coupled $^{31}\mathrm{P}$ electron spins in silicon},\ }\href
  {https://doi.org/10.1103/PhysRevLett.112.236801} {\bibfield  {journal}
  {\bibinfo  {journal} {Phys. Rev. Lett.}\ }\textbf {\bibinfo {volume} {112}},\
  \bibinfo {pages} {236801} (\bibinfo {year} {2014})}\BibitemShut {NoStop}%
\bibitem [{\citenamefont {Watson}\ \emph {et~al.}(2015)\citenamefont {Watson},
  \citenamefont {Weber}, \citenamefont {House}, \citenamefont {B\"uch},\ and\
  \citenamefont {Simmons}}]{watson2015}%
  \BibitemOpen
  \bibfield  {author} {\bibinfo {author} {\bibfnamefont {T.~F.}\ \bibnamefont
  {Watson}}, \bibinfo {author} {\bibfnamefont {B.}~\bibnamefont {Weber}},
  \bibinfo {author} {\bibfnamefont {M.~G.}\ \bibnamefont {House}}, \bibinfo
  {author} {\bibfnamefont {H.}~\bibnamefont {B\"uch}},\ and\ \bibinfo {author}
  {\bibfnamefont {M.~Y.}\ \bibnamefont {Simmons}},\ }\bibfield  {title}
  {\bibinfo {title} {High-fidelity rapid initialization and read-out of an
  electron spin via the single donor ${D}^{\ensuremath{-}}$ charge state},\
  }\href {https://doi.org/10.1103/PhysRevLett.115.166806} {\bibfield  {journal}
  {\bibinfo  {journal} {Phys. Rev. Lett.}\ }\textbf {\bibinfo {volume} {115}},\
  \bibinfo {pages} {166806} (\bibinfo {year} {2015})}\BibitemShut {NoStop}%
\bibitem [{\citenamefont {Broome}\ \emph {et~al.}(2017)\citenamefont {Broome},
  \citenamefont {Watson}, \citenamefont {Keith}, \citenamefont {Gorman},
  \citenamefont {House}, \citenamefont {Keizer}, \citenamefont {Hile},
  \citenamefont {Baker},\ and\ \citenamefont {Simmons}}]{broome2017}%
  \BibitemOpen
  \bibfield  {author} {\bibinfo {author} {\bibfnamefont {M.~A.}\ \bibnamefont
  {Broome}}, \bibinfo {author} {\bibfnamefont {T.~F.}\ \bibnamefont {Watson}},
  \bibinfo {author} {\bibfnamefont {D.}~\bibnamefont {Keith}}, \bibinfo
  {author} {\bibfnamefont {S.~K.}\ \bibnamefont {Gorman}}, \bibinfo {author}
  {\bibfnamefont {M.~G.}\ \bibnamefont {House}}, \bibinfo {author}
  {\bibfnamefont {J.~G.}\ \bibnamefont {Keizer}}, \bibinfo {author}
  {\bibfnamefont {S.~J.}\ \bibnamefont {Hile}}, \bibinfo {author}
  {\bibfnamefont {W.}~\bibnamefont {Baker}},\ and\ \bibinfo {author}
  {\bibfnamefont {M.~Y.}\ \bibnamefont {Simmons}},\ }\bibfield  {title}
  {\bibinfo {title} {High-fidelity single-shot singlet-triplet readout of
  precision-placed donors in silicon},\ }\href
  {https://doi.org/10.1103/PhysRevLett.119.046802} {\bibfield  {journal}
  {\bibinfo  {journal} {Phys. Rev. Lett.}\ }\textbf {\bibinfo {volume} {119}},\
  \bibinfo {pages} {046802} (\bibinfo {year} {2017})}\BibitemShut {NoStop}%
\bibitem [{\citenamefont {Keith}\ \emph {et~al.}(2019)\citenamefont {Keith},
  \citenamefont {House}, \citenamefont {Donnelly}, \citenamefont {Watson},
  \citenamefont {Weber},\ and\ \citenamefont {Simmons}}]{keith2019}%
  \BibitemOpen
  \bibfield  {author} {\bibinfo {author} {\bibfnamefont {D.}~\bibnamefont
  {Keith}}, \bibinfo {author} {\bibfnamefont {M.~G.}\ \bibnamefont {House}},
  \bibinfo {author} {\bibfnamefont {M.~B.}\ \bibnamefont {Donnelly}}, \bibinfo
  {author} {\bibfnamefont {T.~F.}\ \bibnamefont {Watson}}, \bibinfo {author}
  {\bibfnamefont {B.}~\bibnamefont {Weber}},\ and\ \bibinfo {author}
  {\bibfnamefont {M.~Y.}\ \bibnamefont {Simmons}},\ }\bibfield  {title}
  {\bibinfo {title} {Single-shot spin readout in semiconductors near the
  shot-noise sensitivity limit},\ }\href
  {https://doi.org/10.1103/PhysRevX.9.041003} {\bibfield  {journal} {\bibinfo
  {journal} {Phys. Rev. X}\ }\textbf {\bibinfo {volume} {9}},\ \bibinfo {pages}
  {041003} (\bibinfo {year} {2019})}\BibitemShut {NoStop}%
\bibitem [{\citenamefont {Chanrion}\ \emph {et~al.}(2020)\citenamefont
  {Chanrion}, \citenamefont {Niegemann}, \citenamefont {Bertrand},
  \citenamefont {Spence}, \citenamefont {Jadot}, \citenamefont {Li},
  \citenamefont {Mortemousque}, \citenamefont {Hutin}, \citenamefont {Maurand},
  \citenamefont {Jehl}, \citenamefont {Sanquer}, \citenamefont {De~Franceschi},
  \citenamefont {Bäuerle}, \citenamefont {Balestro}, \citenamefont {Niquet},
  \citenamefont {Vinet}, \citenamefont {Meunier},\ and\ \citenamefont
  {Urdampilleta}}]{chanrion2020}%
  \BibitemOpen
  \bibfield  {author} {\bibinfo {author} {\bibfnamefont {E.}~\bibnamefont
  {Chanrion}}, \bibinfo {author} {\bibfnamefont {D.~J.}\ \bibnamefont
  {Niegemann}}, \bibinfo {author} {\bibfnamefont {B.}~\bibnamefont {Bertrand}},
  \bibinfo {author} {\bibfnamefont {C.}~\bibnamefont {Spence}}, \bibinfo
  {author} {\bibfnamefont {B.}~\bibnamefont {Jadot}}, \bibinfo {author}
  {\bibfnamefont {J.}~\bibnamefont {Li}}, \bibinfo {author} {\bibfnamefont
  {P.-A.}\ \bibnamefont {Mortemousque}}, \bibinfo {author} {\bibfnamefont
  {L.}~\bibnamefont {Hutin}}, \bibinfo {author} {\bibfnamefont
  {R.}~\bibnamefont {Maurand}}, \bibinfo {author} {\bibfnamefont
  {X.}~\bibnamefont {Jehl}}, \bibinfo {author} {\bibfnamefont {M.}~\bibnamefont
  {Sanquer}}, \bibinfo {author} {\bibfnamefont {S.}~\bibnamefont
  {De~Franceschi}}, \bibinfo {author} {\bibfnamefont {C.}~\bibnamefont
  {Bäuerle}}, \bibinfo {author} {\bibfnamefont {F.}~\bibnamefont {Balestro}},
  \bibinfo {author} {\bibfnamefont {Y.-M.}\ \bibnamefont {Niquet}}, \bibinfo
  {author} {\bibfnamefont {M.}~\bibnamefont {Vinet}}, \bibinfo {author}
  {\bibfnamefont {T.}~\bibnamefont {Meunier}},\ and\ \bibinfo {author}
  {\bibfnamefont {M.}~\bibnamefont {Urdampilleta}},\ }\bibfield  {title}
  {\bibinfo {title} {Charge detection in an array of {CMOS} quantum dots},\
  }\href {https://doi.org/10.1103/PhysRevApplied.14.024066} {\bibfield
  {journal} {\bibinfo  {journal} {Phys. Rev. Appl.}\ }\textbf {\bibinfo
  {volume} {14}},\ \bibinfo {pages} {024066} (\bibinfo {year}
  {2020})}\BibitemShut {NoStop}%
\bibitem [{\citenamefont {Seedhouse}\ \emph {et~al.}(2021)\citenamefont
  {Seedhouse}, \citenamefont {Tanttu}, \citenamefont {Leon}, \citenamefont
  {Zhao}, \citenamefont {Tan}, \citenamefont {Hensen}, \citenamefont {Hudson},
  \citenamefont {Itoh}, \citenamefont {Yoneda}, \citenamefont {Yang} \emph
  {et~al.}}]{seedhouse2021}%
  \BibitemOpen
  \bibfield  {author} {\bibinfo {author} {\bibfnamefont {A.~E.}\ \bibnamefont
  {Seedhouse}}, \bibinfo {author} {\bibfnamefont {T.}~\bibnamefont {Tanttu}},
  \bibinfo {author} {\bibfnamefont {R.~C.}\ \bibnamefont {Leon}}, \bibinfo
  {author} {\bibfnamefont {R.}~\bibnamefont {Zhao}}, \bibinfo {author}
  {\bibfnamefont {K.~Y.}\ \bibnamefont {Tan}}, \bibinfo {author} {\bibfnamefont
  {B.}~\bibnamefont {Hensen}}, \bibinfo {author} {\bibfnamefont {F.~E.}\
  \bibnamefont {Hudson}}, \bibinfo {author} {\bibfnamefont {K.~M.}\
  \bibnamefont {Itoh}}, \bibinfo {author} {\bibfnamefont {J.}~\bibnamefont
  {Yoneda}}, \bibinfo {author} {\bibfnamefont {C.~H.}\ \bibnamefont {Yang}},
  \emph {et~al.},\ }\bibfield  {title} {\bibinfo {title} {Pauli blockade in
  silicon quantum dots with spin-orbit control},\ }\href@noop {} {\bibfield
  {journal} {\bibinfo  {journal} {PRX Quantum}\ }\textbf {\bibinfo {volume}
  {2}},\ \bibinfo {pages} {010303} (\bibinfo {year} {2021})}\BibitemShut
  {NoStop}%
\bibitem [{\citenamefont {Hofmann}\ \emph {et~al.}(1995)\citenamefont
  {Hofmann}, \citenamefont {Heinzel}, \citenamefont {Wharam}, \citenamefont
  {Kotthaus}, \citenamefont {B\"ohm}, \citenamefont {Klein}, \citenamefont
  {Tr\"ankle},\ and\ \citenamefont {Weimann}}]{hofmann1995}%
  \BibitemOpen
  \bibfield  {author} {\bibinfo {author} {\bibfnamefont {F.}~\bibnamefont
  {Hofmann}}, \bibinfo {author} {\bibfnamefont {T.}~\bibnamefont {Heinzel}},
  \bibinfo {author} {\bibfnamefont {D.}~\bibnamefont {Wharam}}, \bibinfo
  {author} {\bibfnamefont {J.}~\bibnamefont {Kotthaus}}, \bibinfo {author}
  {\bibfnamefont {G.}~\bibnamefont {B\"ohm}}, \bibinfo {author} {\bibfnamefont
  {W.}~\bibnamefont {Klein}}, \bibinfo {author} {\bibfnamefont
  {G.}~\bibnamefont {Tr\"ankle}},\ and\ \bibinfo {author} {\bibfnamefont
  {G.}~\bibnamefont {Weimann}},\ }\bibfield  {title} {\bibinfo {title} {Single
  electron switching in a parallel quantum dot},\ }\href
  {https://doi.org/10.1103/PhysRevB.51.13872} {\bibfield  {journal} {\bibinfo
  {journal} {Phys. Rev. B}\ }\textbf {\bibinfo {volume} {51}},\ \bibinfo
  {pages} {13872} (\bibinfo {year} {1995})}\BibitemShut {NoStop}%
\bibitem [{\citenamefont {Onac}\ \emph {et~al.}(2006)\citenamefont {Onac},
  \citenamefont {Balestro}, \citenamefont {van Beveren}, \citenamefont
  {Hartmann}, \citenamefont {Nazarov},\ and\ \citenamefont
  {Kouwenhoven}}]{onac2006}%
  \BibitemOpen
  \bibfield  {author} {\bibinfo {author} {\bibfnamefont {E.}~\bibnamefont
  {Onac}}, \bibinfo {author} {\bibfnamefont {F.}~\bibnamefont {Balestro}},
  \bibinfo {author} {\bibfnamefont {L.~H.~W.}\ \bibnamefont {van Beveren}},
  \bibinfo {author} {\bibfnamefont {U.}~\bibnamefont {Hartmann}}, \bibinfo
  {author} {\bibfnamefont {Y.~V.}\ \bibnamefont {Nazarov}},\ and\ \bibinfo
  {author} {\bibfnamefont {L.~P.}\ \bibnamefont {Kouwenhoven}},\ }\bibfield
  {title} {\bibinfo {title} {Using a quantum dot as a high-frequency shot noise
  detector},\ }\href {https://doi.org/10.1103/PhysRevLett.96.176601} {\bibfield
   {journal} {\bibinfo  {journal} {Phys. Rev. Lett.}\ }\textbf {\bibinfo
  {volume} {96}},\ \bibinfo {pages} {176601} (\bibinfo {year}
  {2006})}\BibitemShut {NoStop}%
\bibitem [{\citenamefont {Nordberg}\ \emph {et~al.}(2009)\citenamefont
  {Nordberg}, \citenamefont {Stalford}, \citenamefont {Young}, \citenamefont
  {Ten~Eyck}, \citenamefont {Eng}, \citenamefont {Tracy}, \citenamefont
  {Childs}, \citenamefont {Wendt}, \citenamefont {Grubbs}, \citenamefont
  {Stevens}, \citenamefont {Lilly}, \citenamefont {Eriksson},\ and\
  \citenamefont {Carroll}}]{nordberg2009}%
  \BibitemOpen
  \bibfield  {author} {\bibinfo {author} {\bibfnamefont {E.~P.}\ \bibnamefont
  {Nordberg}}, \bibinfo {author} {\bibfnamefont {H.~L.}\ \bibnamefont
  {Stalford}}, \bibinfo {author} {\bibfnamefont {R.}~\bibnamefont {Young}},
  \bibinfo {author} {\bibfnamefont {G.~A.}\ \bibnamefont {Ten~Eyck}}, \bibinfo
  {author} {\bibfnamefont {K.}~\bibnamefont {Eng}}, \bibinfo {author}
  {\bibfnamefont {L.~A.}\ \bibnamefont {Tracy}}, \bibinfo {author}
  {\bibfnamefont {K.~D.}\ \bibnamefont {Childs}}, \bibinfo {author}
  {\bibfnamefont {J.~R.}\ \bibnamefont {Wendt}}, \bibinfo {author}
  {\bibfnamefont {R.~K.}\ \bibnamefont {Grubbs}}, \bibinfo {author}
  {\bibfnamefont {J.}~\bibnamefont {Stevens}}, \bibinfo {author} {\bibfnamefont
  {M.~P.}\ \bibnamefont {Lilly}}, \bibinfo {author} {\bibfnamefont {M.~A.}\
  \bibnamefont {Eriksson}},\ and\ \bibinfo {author} {\bibfnamefont {M.~S.}\
  \bibnamefont {Carroll}},\ }\bibfield  {title} {\bibinfo {title} {Charge
  sensing in enhancement mode double-top-gated metal-oxide-semiconductor
  quantum dots},\ }\href {https://doi.org/10.1063/1.3259416} {\bibfield
  {journal} {\bibinfo  {journal} {Appl. Phys. Lett.}\ }\textbf {\bibinfo
  {volume} {95}},\ \bibinfo {pages} {202102} (\bibinfo {year}
  {2009})}\BibitemShut {NoStop}%
\bibitem [{\citenamefont {Reilly}\ \emph {et~al.}(2007)\citenamefont {Reilly},
  \citenamefont {Marcus}, \citenamefont {Hanson},\ and\ \citenamefont
  {Gossard}}]{reilly2007}%
  \BibitemOpen
  \bibfield  {author} {\bibinfo {author} {\bibfnamefont {D.~J.}\ \bibnamefont
  {Reilly}}, \bibinfo {author} {\bibfnamefont {C.~M.}\ \bibnamefont {Marcus}},
  \bibinfo {author} {\bibfnamefont {M.~P.}\ \bibnamefont {Hanson}},\ and\
  \bibinfo {author} {\bibfnamefont {A.~C.}\ \bibnamefont {Gossard}},\
  }\bibfield  {title} {\bibinfo {title} {Fast single-charge sensing with a {RF}
  quantum point contact},\ }\href
  {https://doi.org/http://dx.doi.org/10.1063/1.2794995} {\bibfield  {journal}
  {\bibinfo  {journal} {Appl. Phys. Lett.}\ }\textbf {\bibinfo {volume} {91}},\
  \bibinfo {eid} {162101} (\bibinfo {year} {2007})}\BibitemShut {NoStop}%
\bibitem [{\citenamefont {Barthel}\ \emph {et~al.}(2010)\citenamefont
  {Barthel}, \citenamefont {Kj\ae{}rgaard}, \citenamefont {Medford},
  \citenamefont {Stopa}, \citenamefont {Marcus}, \citenamefont {Hanson},\ and\
  \citenamefont {Gossard}}]{barthel2010}%
  \BibitemOpen
  \bibfield  {author} {\bibinfo {author} {\bibfnamefont {C.}~\bibnamefont
  {Barthel}}, \bibinfo {author} {\bibfnamefont {M.}~\bibnamefont
  {Kj\ae{}rgaard}}, \bibinfo {author} {\bibfnamefont {J.}~\bibnamefont
  {Medford}}, \bibinfo {author} {\bibfnamefont {M.}~\bibnamefont {Stopa}},
  \bibinfo {author} {\bibfnamefont {C.~M.}\ \bibnamefont {Marcus}}, \bibinfo
  {author} {\bibfnamefont {M.~P.}\ \bibnamefont {Hanson}},\ and\ \bibinfo
  {author} {\bibfnamefont {A.~C.}\ \bibnamefont {Gossard}},\ }\bibfield
  {title} {\bibinfo {title} {Fast sensing of double-dot charge arrangement and
  spin state with a radio-frequency sensor quantum dot},\ }\href
  {https://doi.org/10.1103/PhysRevB.81.161308} {\bibfield  {journal} {\bibinfo
  {journal} {Phys. Rev. B}\ }\textbf {\bibinfo {volume} {81}},\ \bibinfo
  {pages} {161308} (\bibinfo {year} {2010})}\BibitemShut {NoStop}%
\bibitem [{\citenamefont {Schoelkopf}\ \emph {et~al.}(1998)\citenamefont
  {Schoelkopf}, \citenamefont {Wahlgren}, \citenamefont {Kozhevnikov},
  \citenamefont {Delsing},\ and\ \citenamefont {Prober}}]{schoelkopf1998}%
  \BibitemOpen
  \bibfield  {author} {\bibinfo {author} {\bibfnamefont {R.~J.}\ \bibnamefont
  {Schoelkopf}}, \bibinfo {author} {\bibfnamefont {P.}~\bibnamefont
  {Wahlgren}}, \bibinfo {author} {\bibfnamefont {A.~A.}\ \bibnamefont
  {Kozhevnikov}}, \bibinfo {author} {\bibfnamefont {P.}~\bibnamefont
  {Delsing}},\ and\ \bibinfo {author} {\bibfnamefont {D.~E.}\ \bibnamefont
  {Prober}},\ }\bibfield  {title} {\bibinfo {title} {The radio-frequency
  single-electron transistor {(RF-SET)}: A fast and ultrasensitive
  electrometer},\ }\href {https://doi.org/10.1126/science.280.5367.1238}
  {\bibfield  {journal} {\bibinfo  {journal} {Science}\ }\textbf {\bibinfo
  {volume} {280}},\ \bibinfo {pages} {1238} (\bibinfo {year}
  {1998})}\BibitemShut {NoStop}%
\bibitem [{\citenamefont {Petersson}\ \emph {et~al.}(2010)\citenamefont
  {Petersson}, \citenamefont {Smith}, \citenamefont {Anderson}, \citenamefont
  {Atkinson}, \citenamefont {Jones},\ and\ \citenamefont
  {Ritchie}}]{petersson2010}%
  \BibitemOpen
  \bibfield  {author} {\bibinfo {author} {\bibfnamefont {K.~D.}\ \bibnamefont
  {Petersson}}, \bibinfo {author} {\bibfnamefont {C.~G.}\ \bibnamefont
  {Smith}}, \bibinfo {author} {\bibfnamefont {D.}~\bibnamefont {Anderson}},
  \bibinfo {author} {\bibfnamefont {P.}~\bibnamefont {Atkinson}}, \bibinfo
  {author} {\bibfnamefont {G.~A.~C.}\ \bibnamefont {Jones}},\ and\ \bibinfo
  {author} {\bibfnamefont {D.~A.}\ \bibnamefont {Ritchie}},\ }\bibfield
  {title} {\bibinfo {title} {Charge and spin state readout of a double quantum
  dot coupled to a resonator},\ }\href {https://doi.org/10.1021/nl100663w}
  {\bibfield  {journal} {\bibinfo  {journal} {Nano Letters}\ }\textbf {\bibinfo
  {volume} {10}},\ \bibinfo {pages} {2789} (\bibinfo {year}
  {2010})}\BibitemShut {NoStop}%
\bibitem [{\citenamefont {Petersson}\ \emph {et~al.}(2012)\citenamefont
  {Petersson}, \citenamefont {McFaul}, \citenamefont {Schroer}, \citenamefont
  {Jung}, \citenamefont {Taylor}, \citenamefont {Houck},\ and\ \citenamefont
  {Petta}}]{petersson2012}%
  \BibitemOpen
  \bibfield  {author} {\bibinfo {author} {\bibfnamefont {K.}~\bibnamefont
  {Petersson}}, \bibinfo {author} {\bibfnamefont {L.}~\bibnamefont {McFaul}},
  \bibinfo {author} {\bibfnamefont {M.}~\bibnamefont {Schroer}}, \bibinfo
  {author} {\bibfnamefont {M.}~\bibnamefont {Jung}}, \bibinfo {author}
  {\bibfnamefont {J.}~\bibnamefont {Taylor}}, \bibinfo {author} {\bibfnamefont
  {A.}~\bibnamefont {Houck}},\ and\ \bibinfo {author} {\bibfnamefont
  {J.}~\bibnamefont {Petta}},\ }\bibfield  {title} {\bibinfo {title} {Circuit
  quantum electrodynamics with a spin qubit},\ }\href
  {https://doi.org/doi:10.1038/nature11559} {\bibfield  {journal} {\bibinfo
  {journal} {Nature}\ }\textbf {\bibinfo {volume} {490}},\ \bibinfo {pages}
  {380} (\bibinfo {year} {2012})}\BibitemShut {NoStop}%
\bibitem [{\citenamefont {Stehlik}\ \emph {et~al.}(2015)\citenamefont
  {Stehlik}, \citenamefont {Liu}, \citenamefont {Quintana}, \citenamefont
  {Eichler}, \citenamefont {Hartke},\ and\ \citenamefont
  {Petta}}]{stehlik2015}%
  \BibitemOpen
  \bibfield  {author} {\bibinfo {author} {\bibfnamefont {J.}~\bibnamefont
  {Stehlik}}, \bibinfo {author} {\bibfnamefont {Y.-Y.}\ \bibnamefont {Liu}},
  \bibinfo {author} {\bibfnamefont {C.~M.}\ \bibnamefont {Quintana}}, \bibinfo
  {author} {\bibfnamefont {C.}~\bibnamefont {Eichler}}, \bibinfo {author}
  {\bibfnamefont {T.~R.}\ \bibnamefont {Hartke}},\ and\ \bibinfo {author}
  {\bibfnamefont {J.~R.}\ \bibnamefont {Petta}},\ }\bibfield  {title} {\bibinfo
  {title} {Fast charge sensing of a cavity-coupled double quantum dot using a
  {Josephson} parametric amplifier},\ }\href
  {https://doi.org/10.1103/PhysRevApplied.4.014018} {\bibfield  {journal}
  {\bibinfo  {journal} {Phys. Rev. Appl.}\ }\textbf {\bibinfo {volume} {4}},\
  \bibinfo {pages} {014018} (\bibinfo {year} {2015})}\BibitemShut {NoStop}%
\bibitem [{\citenamefont {Crippa}\ \emph {et~al.}(2019)\citenamefont {Crippa},
  \citenamefont {Ezzouch}, \citenamefont {Aprá}, \citenamefont {Amisse},
  \citenamefont {Laviéville}, \citenamefont {Hutin}, \citenamefont {Bertrand},
  \citenamefont {Vinet}, \citenamefont {Urdampilleta}, \citenamefont {Meunier},
  \citenamefont {Sanquer}, \citenamefont {Jehl}, \citenamefont {Maurand},\ and\
  \citenamefont {De~Franceschi}}]{crippa2019}%
  \BibitemOpen
  \bibfield  {author} {\bibinfo {author} {\bibfnamefont {A.}~\bibnamefont
  {Crippa}}, \bibinfo {author} {\bibfnamefont {R.}~\bibnamefont {Ezzouch}},
  \bibinfo {author} {\bibfnamefont {A.}~\bibnamefont {Aprá}}, \bibinfo
  {author} {\bibfnamefont {A.}~\bibnamefont {Amisse}}, \bibinfo {author}
  {\bibfnamefont {R.}~\bibnamefont {Laviéville}}, \bibinfo {author}
  {\bibfnamefont {L.}~\bibnamefont {Hutin}}, \bibinfo {author} {\bibfnamefont
  {B.}~\bibnamefont {Bertrand}}, \bibinfo {author} {\bibfnamefont
  {M.}~\bibnamefont {Vinet}}, \bibinfo {author} {\bibfnamefont
  {M.}~\bibnamefont {Urdampilleta}}, \bibinfo {author} {\bibfnamefont
  {T.}~\bibnamefont {Meunier}}, \bibinfo {author} {\bibfnamefont
  {M.}~\bibnamefont {Sanquer}}, \bibinfo {author} {\bibfnamefont
  {X.}~\bibnamefont {Jehl}}, \bibinfo {author} {\bibfnamefont {R.}~\bibnamefont
  {Maurand}},\ and\ \bibinfo {author} {\bibfnamefont {S.}~\bibnamefont
  {De~Franceschi}},\ }\bibfield  {title} {\bibinfo {title} {Gate-reflectometry
  dispersive readout and coherent control of a spin qubit in silicon},\ }\href
  {https://doi.org/10.1038/s41467-019-10848-z} {\bibfield  {journal} {\bibinfo
  {journal} {Nat. Commun.}\ }\textbf {\bibinfo {volume} {10}},\ \bibinfo
  {pages} {2776} (\bibinfo {year} {2019})}\BibitemShut {NoStop}%
\bibitem [{\citenamefont {Urdampilleta}\ \emph {et~al.}(2019)\citenamefont
  {Urdampilleta}, \citenamefont {Niegemann}, \citenamefont {Chanrion},
  \citenamefont {Jadot}, \citenamefont {Spence}, \citenamefont {Mortemousque},
  \citenamefont {Bäuerle}, \citenamefont {Hutin}, \citenamefont {Bertrand},
  \citenamefont {Barraud}, \citenamefont {Maurand}, \citenamefont {Sanquer},
  \citenamefont {Jehl}, \citenamefont {De~Franceschi}, \citenamefont {Vinet},\
  and\ \citenamefont {Meunier}}]{urdampilleta2019}%
  \BibitemOpen
  \bibfield  {author} {\bibinfo {author} {\bibfnamefont {M.}~\bibnamefont
  {Urdampilleta}}, \bibinfo {author} {\bibfnamefont {D.~J.}\ \bibnamefont
  {Niegemann}}, \bibinfo {author} {\bibfnamefont {E.}~\bibnamefont {Chanrion}},
  \bibinfo {author} {\bibfnamefont {B.}~\bibnamefont {Jadot}}, \bibinfo
  {author} {\bibfnamefont {C.}~\bibnamefont {Spence}}, \bibinfo {author}
  {\bibfnamefont {P.-A.}\ \bibnamefont {Mortemousque}}, \bibinfo {author}
  {\bibfnamefont {C.}~\bibnamefont {Bäuerle}}, \bibinfo {author}
  {\bibfnamefont {L.}~\bibnamefont {Hutin}}, \bibinfo {author} {\bibfnamefont
  {B.}~\bibnamefont {Bertrand}}, \bibinfo {author} {\bibfnamefont
  {S.}~\bibnamefont {Barraud}}, \bibinfo {author} {\bibfnamefont
  {R.}~\bibnamefont {Maurand}}, \bibinfo {author} {\bibfnamefont
  {M.}~\bibnamefont {Sanquer}}, \bibinfo {author} {\bibfnamefont
  {X.}~\bibnamefont {Jehl}}, \bibinfo {author} {\bibfnamefont {S.}~\bibnamefont
  {De~Franceschi}}, \bibinfo {author} {\bibfnamefont {M.}~\bibnamefont
  {Vinet}},\ and\ \bibinfo {author} {\bibfnamefont {T.}~\bibnamefont
  {Meunier}},\ }\bibfield  {title} {\bibinfo {title} {Gate-based high fidelity
  spin readout in a {CMOS} device},\ }\href
  {https://doi.org/10.1038/s41565-019-0443-9} {\bibfield  {journal} {\bibinfo
  {journal} {Nat. Nanotechnol.}\ }\textbf {\bibinfo {volume} {14}},\ \bibinfo
  {pages} {737} (\bibinfo {year} {2019})}\BibitemShut {NoStop}%
\bibitem [{\citenamefont {Ciriano-Tejel}\ \emph {et~al.}(2021)\citenamefont
  {Ciriano-Tejel}, \citenamefont {Fogarty}, \citenamefont {Schaal},
  \citenamefont {Hutin}, \citenamefont {Bertrand}, \citenamefont {Ibberson},
  \citenamefont {Gonzalez-Zalba}, \citenamefont {Li}, \citenamefont {Niquet},
  \citenamefont {Vinet},\ and\ \citenamefont {Morton}}]{ciriano-tejel2021}%
  \BibitemOpen
  \bibfield  {author} {\bibinfo {author} {\bibfnamefont {V.~N.}\ \bibnamefont
  {Ciriano-Tejel}}, \bibinfo {author} {\bibfnamefont {M.~A.}\ \bibnamefont
  {Fogarty}}, \bibinfo {author} {\bibfnamefont {S.}~\bibnamefont {Schaal}},
  \bibinfo {author} {\bibfnamefont {L.}~\bibnamefont {Hutin}}, \bibinfo
  {author} {\bibfnamefont {B.}~\bibnamefont {Bertrand}}, \bibinfo {author}
  {\bibfnamefont {L.}~\bibnamefont {Ibberson}}, \bibinfo {author}
  {\bibfnamefont {M.~F.}\ \bibnamefont {Gonzalez-Zalba}}, \bibinfo {author}
  {\bibfnamefont {J.}~\bibnamefont {Li}}, \bibinfo {author} {\bibfnamefont
  {Y.-M.}\ \bibnamefont {Niquet}}, \bibinfo {author} {\bibfnamefont
  {M.}~\bibnamefont {Vinet}},\ and\ \bibinfo {author} {\bibfnamefont {J.~J.}\
  \bibnamefont {Morton}},\ }\bibfield  {title} {\bibinfo {title} {Spin readout
  of a {CMOS} quantum dot by gate reflectometry and spin-dependent tunneling},\
  }\href {https://doi.org/10.1103/PRXQuantum.2.010353} {\bibfield  {journal}
  {\bibinfo  {journal} {PRX Quantum}\ }\textbf {\bibinfo {volume} {2}},\
  \bibinfo {pages} {010353} (\bibinfo {year} {2021})}\BibitemShut {NoStop}%
\bibitem [{\citenamefont {Ansaloni}\ \emph {et~al.}(2020)\citenamefont
  {Ansaloni}, \citenamefont {Chatterjee}, \citenamefont {Bohuslavskyi},
  \citenamefont {Bertrand}, \citenamefont {Hutin}, \citenamefont {Vinet},\ and\
  \citenamefont {Kuemmeth}}]{ansaloni2020}%
  \BibitemOpen
  \bibfield  {author} {\bibinfo {author} {\bibfnamefont {F.}~\bibnamefont
  {Ansaloni}}, \bibinfo {author} {\bibfnamefont {A.}~\bibnamefont
  {Chatterjee}}, \bibinfo {author} {\bibfnamefont {H.}~\bibnamefont
  {Bohuslavskyi}}, \bibinfo {author} {\bibfnamefont {B.}~\bibnamefont
  {Bertrand}}, \bibinfo {author} {\bibfnamefont {L.}~\bibnamefont {Hutin}},
  \bibinfo {author} {\bibfnamefont {M.}~\bibnamefont {Vinet}},\ and\ \bibinfo
  {author} {\bibfnamefont {F.}~\bibnamefont {Kuemmeth}},\ }\bibfield  {title}
  {\bibinfo {title} {Single-electron operations in a foundry-fabricated array
  of quantum dots},\ }\href {https://doi.org/10.1038/s41467-020-20280-3}
  {\bibfield  {journal} {\bibinfo  {journal} {Nat. Commun.}\ }\textbf {\bibinfo
  {volume} {11}},\ \bibinfo {pages} {6399} (\bibinfo {year}
  {2020})}\BibitemShut {NoStop}%
\bibitem [{\citenamefont {Friesen}\ \emph {et~al.}(2004)\citenamefont
  {Friesen}, \citenamefont {Tahan}, \citenamefont {Joynt},\ and\ \citenamefont
  {Eriksson}}]{friesen2004}%
  \BibitemOpen
  \bibfield  {author} {\bibinfo {author} {\bibfnamefont {M.}~\bibnamefont
  {Friesen}}, \bibinfo {author} {\bibfnamefont {C.}~\bibnamefont {Tahan}},
  \bibinfo {author} {\bibfnamefont {R.}~\bibnamefont {Joynt}},\ and\ \bibinfo
  {author} {\bibfnamefont {M.~A.}\ \bibnamefont {Eriksson}},\ }\bibfield
  {title} {\bibinfo {title} {Spin readout and initialization in a semiconductor
  quantum dot},\ }\href {https://doi.org/10.1103/PhysRevLett.92.037901}
  {\bibfield  {journal} {\bibinfo  {journal} {Phys. Rev. Lett.}\ }\textbf
  {\bibinfo {volume} {92}},\ \bibinfo {pages} {037901} (\bibinfo {year}
  {2004})}\BibitemShut {NoStop}%
\bibitem [{\citenamefont {Petta}\ \emph {et~al.}(2005)\citenamefont {Petta},
  \citenamefont {Johnson}, \citenamefont {Taylor}, \citenamefont {Laird},
  \citenamefont {Yacoby}, \citenamefont {Lukin}, \citenamefont {Marcus},
  \citenamefont {Hanson},\ and\ \citenamefont {Gossard}}]{petta2005}%
  \BibitemOpen
  \bibfield  {author} {\bibinfo {author} {\bibfnamefont {J.~R.}\ \bibnamefont
  {Petta}}, \bibinfo {author} {\bibfnamefont {A.~C.}\ \bibnamefont {Johnson}},
  \bibinfo {author} {\bibfnamefont {J.~M.}\ \bibnamefont {Taylor}}, \bibinfo
  {author} {\bibfnamefont {E.~A.}\ \bibnamefont {Laird}}, \bibinfo {author}
  {\bibfnamefont {A.}~\bibnamefont {Yacoby}}, \bibinfo {author} {\bibfnamefont
  {M.~D.}\ \bibnamefont {Lukin}}, \bibinfo {author} {\bibfnamefont {C.~M.}\
  \bibnamefont {Marcus}}, \bibinfo {author} {\bibfnamefont {M.~P.}\
  \bibnamefont {Hanson}},\ and\ \bibinfo {author} {\bibfnamefont {A.~C.}\
  \bibnamefont {Gossard}},\ }\bibfield  {title} {\bibinfo {title} {Coherent
  manipulation of coupled electron spins in semiconductor quantum dots},\
  }\href {https://doi.org/10.1126/science.1116955} {\bibfield  {journal}
  {\bibinfo  {journal} {Science}\ }\textbf {\bibinfo {volume} {309}},\ \bibinfo
  {pages} {2180} (\bibinfo {year} {2005})},\ \Eprint
  {https://arxiv.org/abs/http://science.sciencemag.org/content/309/5744/2180.full.pdf}
  {http://science.sciencemag.org/content/309/5744/2180.full.pdf} \BibitemShut
  {NoStop}%
\bibitem [{\citenamefont {Nakajima}\ \emph {et~al.}(2019)\citenamefont
  {Nakajima}, \citenamefont {Noiri}, \citenamefont {Yoneda}, \citenamefont
  {Delbecq}, \citenamefont {Stano}, \citenamefont {Otsuka}, \citenamefont
  {Takeda}, \citenamefont {Amaha}, \citenamefont {Allison}, \citenamefont
  {Kawasaki} \emph {et~al.}}]{nakajima2019}%
  \BibitemOpen
  \bibfield  {author} {\bibinfo {author} {\bibfnamefont {T.}~\bibnamefont
  {Nakajima}}, \bibinfo {author} {\bibfnamefont {A.}~\bibnamefont {Noiri}},
  \bibinfo {author} {\bibfnamefont {J.}~\bibnamefont {Yoneda}}, \bibinfo
  {author} {\bibfnamefont {M.~R.}\ \bibnamefont {Delbecq}}, \bibinfo {author}
  {\bibfnamefont {P.}~\bibnamefont {Stano}}, \bibinfo {author} {\bibfnamefont
  {T.}~\bibnamefont {Otsuka}}, \bibinfo {author} {\bibfnamefont
  {K.}~\bibnamefont {Takeda}}, \bibinfo {author} {\bibfnamefont
  {S.}~\bibnamefont {Amaha}}, \bibinfo {author} {\bibfnamefont
  {G.}~\bibnamefont {Allison}}, \bibinfo {author} {\bibfnamefont
  {K.}~\bibnamefont {Kawasaki}}, \emph {et~al.},\ }\bibfield  {title} {\bibinfo
  {title} {Quantum non-demolition measurement of an electron spin qubit},\
  }\href@noop {} {\bibfield  {journal} {\bibinfo  {journal} {Nat.
  Nanotechnol.}\ }\textbf {\bibinfo {volume} {14}},\ \bibinfo {pages} {555}
  (\bibinfo {year} {2019})}\BibitemShut {NoStop}%
\bibitem [{\citenamefont {Zajac}\ \emph {et~al.}(2015)\citenamefont {Zajac},
  \citenamefont {Hazard}, \citenamefont {Mi}, \citenamefont {Wang},\ and\
  \citenamefont {Petta}}]{Zajac2015}%
  \BibitemOpen
  \bibfield  {author} {\bibinfo {author} {\bibfnamefont {D.~M.}\ \bibnamefont
  {Zajac}}, \bibinfo {author} {\bibfnamefont {T.~M.}\ \bibnamefont {Hazard}},
  \bibinfo {author} {\bibfnamefont {X.}~\bibnamefont {Mi}}, \bibinfo {author}
  {\bibfnamefont {K.}~\bibnamefont {Wang}},\ and\ \bibinfo {author}
  {\bibfnamefont {J.~R.}\ \bibnamefont {Petta}},\ }\bibfield  {title} {\bibinfo
  {title} {A reconfigurable gate architecture for {Si/SiGe} quantum dots},\
  }\href {https://doi.org/10.1063/1.4922249} {\bibfield  {journal} {\bibinfo
  {journal} {Appl. Phys. Lett.}\ }\textbf {\bibinfo {volume} {106}},\ \bibinfo
  {pages} {223507} (\bibinfo {year} {2015})}\BibitemShut {NoStop}%
\bibitem [{\citenamefont {Borselli}\ \emph {et~al.}(2015)\citenamefont
  {Borselli}, \citenamefont {Eng}, \citenamefont {Ross}, \citenamefont
  {Hazard}, \citenamefont {Holabird}, \citenamefont {Huang}, \citenamefont
  {Kiselev}, \citenamefont {Deelman}, \citenamefont {Warren}, \citenamefont
  {Milosavljevic}, \citenamefont {Schmitz}, \citenamefont {Sokolich},
  \citenamefont {Gyure},\ and\ \citenamefont {Hunter}}]{Borselli2015}%
  \BibitemOpen
  \bibfield  {author} {\bibinfo {author} {\bibfnamefont {M.~G.}\ \bibnamefont
  {Borselli}}, \bibinfo {author} {\bibfnamefont {K.}~\bibnamefont {Eng}},
  \bibinfo {author} {\bibfnamefont {R.~S.}\ \bibnamefont {Ross}}, \bibinfo
  {author} {\bibfnamefont {T.~M.}\ \bibnamefont {Hazard}}, \bibinfo {author}
  {\bibfnamefont {K.~S.}\ \bibnamefont {Holabird}}, \bibinfo {author}
  {\bibfnamefont {B.}~\bibnamefont {Huang}}, \bibinfo {author} {\bibfnamefont
  {A.~A.}\ \bibnamefont {Kiselev}}, \bibinfo {author} {\bibfnamefont {P.~W.}\
  \bibnamefont {Deelman}}, \bibinfo {author} {\bibfnamefont {L.~D.}\
  \bibnamefont {Warren}}, \bibinfo {author} {\bibfnamefont {I.}~\bibnamefont
  {Milosavljevic}}, \bibinfo {author} {\bibfnamefont {A.~E.}\ \bibnamefont
  {Schmitz}}, \bibinfo {author} {\bibfnamefont {M.}~\bibnamefont {Sokolich}},
  \bibinfo {author} {\bibfnamefont {M.~F.}\ \bibnamefont {Gyure}},\ and\
  \bibinfo {author} {\bibfnamefont {A.~T.}\ \bibnamefont {Hunter}},\ }\bibfield
   {title} {\bibinfo {title} {Undoped accumulation-mode {Si/SiGe} quantum
  dots},\ }\href {http://stacks.iop.org/0957-4484/26/i=37/a=375202} {\bibfield
  {journal} {\bibinfo  {journal} {Nanotechnology}\ }\textbf {\bibinfo {volume}
  {26}},\ \bibinfo {pages} {375202} (\bibinfo {year} {2015})}\BibitemShut
  {NoStop}%
\bibitem [{\citenamefont {Levy}(2002)}]{levy2002}%
  \BibitemOpen
  \bibfield  {author} {\bibinfo {author} {\bibfnamefont {J.}~\bibnamefont
  {Levy}},\ }\bibfield  {title} {\bibinfo {title} {Universal quantum
  computation with spin-$1/2$ pairs and {Heisenberg} exchange},\ }\href
  {https://doi.org/10.1103/PhysRevLett.89.147902} {\bibfield  {journal}
  {\bibinfo  {journal} {Phys. Rev. Lett.}\ }\textbf {\bibinfo {volume} {89}},\
  \bibinfo {pages} {147902} (\bibinfo {year} {2002})}\BibitemShut {NoStop}%
\bibitem [{\citenamefont {Barthel}\ \emph {et~al.}(2009)\citenamefont
  {Barthel}, \citenamefont {Reilly}, \citenamefont {Marcus}, \citenamefont
  {Hanson},\ and\ \citenamefont {Gossard}}]{barthel2009}%
  \BibitemOpen
  \bibfield  {author} {\bibinfo {author} {\bibfnamefont {C.}~\bibnamefont
  {Barthel}}, \bibinfo {author} {\bibfnamefont {D.~J.}\ \bibnamefont {Reilly}},
  \bibinfo {author} {\bibfnamefont {C.~M.}\ \bibnamefont {Marcus}}, \bibinfo
  {author} {\bibfnamefont {M.~P.}\ \bibnamefont {Hanson}},\ and\ \bibinfo
  {author} {\bibfnamefont {A.~C.}\ \bibnamefont {Gossard}},\ }\bibfield
  {title} {\bibinfo {title} {Rapid single-shot measurement of a singlet-triplet
  qubit},\ }\href {https://doi.org/10.1103/PhysRevLett.103.160503} {\bibfield
  {journal} {\bibinfo  {journal} {Phys. Rev. Lett.}\ }\textbf {\bibinfo
  {volume} {103}},\ \bibinfo {pages} {160503} (\bibinfo {year}
  {2009})}\BibitemShut {NoStop}%
\bibitem [{\citenamefont {Tracy}\ \emph {et~al.}(2016)\citenamefont {Tracy},
  \citenamefont {Luhman}, \citenamefont {Carr}, \citenamefont {Bishop},
  \citenamefont {Ten~Eyck}, \citenamefont {Pluym}, \citenamefont {Wendt},
  \citenamefont {Lilly},\ and\ \citenamefont {Carroll}}]{tracy2016}%
  \BibitemOpen
  \bibfield  {author} {\bibinfo {author} {\bibfnamefont {L.~A.}\ \bibnamefont
  {Tracy}}, \bibinfo {author} {\bibfnamefont {D.~R.}\ \bibnamefont {Luhman}},
  \bibinfo {author} {\bibfnamefont {S.~M.}\ \bibnamefont {Carr}}, \bibinfo
  {author} {\bibfnamefont {N.~C.}\ \bibnamefont {Bishop}}, \bibinfo {author}
  {\bibfnamefont {G.~A.}\ \bibnamefont {Ten~Eyck}}, \bibinfo {author}
  {\bibfnamefont {T.}~\bibnamefont {Pluym}}, \bibinfo {author} {\bibfnamefont
  {J.~R.}\ \bibnamefont {Wendt}}, \bibinfo {author} {\bibfnamefont {M.~P.}\
  \bibnamefont {Lilly}},\ and\ \bibinfo {author} {\bibfnamefont {M.~S.}\
  \bibnamefont {Carroll}},\ }\bibfield  {title} {\bibinfo {title} {Single shot
  spin readout using a cryogenic high-electron-mobility transistor amplifier at
  sub-{Kelvin} temperatures},\ }\href
  {https://doi.org/http://dx.doi.org/10.1063/1.4941421} {\bibfield  {journal}
  {\bibinfo  {journal} {Appl. Phys. Lett.}\ }\textbf {\bibinfo {volume}
  {108}},\ \bibinfo {eid} {063101} (\bibinfo {year} {2016})}\BibitemShut
  {NoStop}%
\bibitem [{\citenamefont {Vink}\ \emph {et~al.}(2007)\citenamefont {Vink},
  \citenamefont {Nooitgedagt}, \citenamefont {Schouten}, \citenamefont
  {Vandersypen},\ and\ \citenamefont {Wegscheider}}]{vink2004}%
  \BibitemOpen
  \bibfield  {author} {\bibinfo {author} {\bibfnamefont {I.~T.}\ \bibnamefont
  {Vink}}, \bibinfo {author} {\bibfnamefont {T.}~\bibnamefont {Nooitgedagt}},
  \bibinfo {author} {\bibfnamefont {R.~N.}\ \bibnamefont {Schouten}}, \bibinfo
  {author} {\bibfnamefont {L.~M.~K.}\ \bibnamefont {Vandersypen}},\ and\
  \bibinfo {author} {\bibfnamefont {W.}~\bibnamefont {Wegscheider}},\
  }\bibfield  {title} {\bibinfo {title} {Cryogenic amplifier for fast real-time
  detection of single-electron tunneling},\ }\href
  {https://doi.org/http://dx.doi.org/10.1063/1.2783265} {\bibfield  {journal}
  {\bibinfo  {journal} {Appl. Phys. Lett.}\ }\textbf {\bibinfo {volume} {91}},\
  \bibinfo {eid} {123512} (\bibinfo {year} {2007})}\BibitemShut {NoStop}%
\bibitem [{\citenamefont {Jones}\ \emph {et~al.}(2019)\citenamefont {Jones},
  \citenamefont {Pritchett}, \citenamefont {Chen}, \citenamefont {Keating},
  \citenamefont {Andrews}, \citenamefont {Blumoff}, \citenamefont {De~Lorenzo},
  \citenamefont {Eng}, \citenamefont {Ha}, \citenamefont {Kiselev},
  \citenamefont {Meenehan}, \citenamefont {Merkel}, \citenamefont {Wright},
  \citenamefont {Edge}, \citenamefont {Ross}, \citenamefont {Rakher},
  \citenamefont {Borselli},\ and\ \citenamefont {Hunter}}]{jones2019}%
  \BibitemOpen
  \bibfield  {author} {\bibinfo {author} {\bibfnamefont {A.}~\bibnamefont
  {Jones}}, \bibinfo {author} {\bibfnamefont {E.}~\bibnamefont {Pritchett}},
  \bibinfo {author} {\bibfnamefont {E.}~\bibnamefont {Chen}}, \bibinfo {author}
  {\bibfnamefont {T.}~\bibnamefont {Keating}}, \bibinfo {author} {\bibfnamefont
  {R.}~\bibnamefont {Andrews}}, \bibinfo {author} {\bibfnamefont
  {J.}~\bibnamefont {Blumoff}}, \bibinfo {author} {\bibfnamefont
  {L.}~\bibnamefont {De~Lorenzo}}, \bibinfo {author} {\bibfnamefont
  {K.}~\bibnamefont {Eng}}, \bibinfo {author} {\bibfnamefont {S.}~\bibnamefont
  {Ha}}, \bibinfo {author} {\bibfnamefont {A.}~\bibnamefont {Kiselev}},
  \bibinfo {author} {\bibfnamefont {S.}~\bibnamefont {Meenehan}}, \bibinfo
  {author} {\bibfnamefont {S.}~\bibnamefont {Merkel}}, \bibinfo {author}
  {\bibfnamefont {J.}~\bibnamefont {Wright}}, \bibinfo {author} {\bibfnamefont
  {L.}~\bibnamefont {Edge}}, \bibinfo {author} {\bibfnamefont {R.}~\bibnamefont
  {Ross}}, \bibinfo {author} {\bibfnamefont {M.}~\bibnamefont {Rakher}},
  \bibinfo {author} {\bibfnamefont {M.}~\bibnamefont {Borselli}},\ and\
  \bibinfo {author} {\bibfnamefont {A.}~\bibnamefont {Hunter}},\ }\bibfield
  {title} {\bibinfo {title} {Spin-blockade spectroscopy of
  $\mathrm{Si}$/$\mathrm{Si}$-$\mathrm{Ge}$ quantum dots},\ }\href
  {https://doi.org/10.1103/PhysRevApplied.12.014026} {\bibfield  {journal}
  {\bibinfo  {journal} {Phys. Rev. Appl.}\ }\textbf {\bibinfo {volume} {12}},\
  \bibinfo {pages} {014026} (\bibinfo {year} {2019})}\BibitemShut {NoStop}%
\bibitem [{\citenamefont {Veldhorst}\ \emph {et~al.}(2014)\citenamefont
  {Veldhorst}, \citenamefont {Hwang}, \citenamefont {Yang}, \citenamefont
  {Leenstra}, \citenamefont {de~Ronde}, \citenamefont {Dehollain},
  \citenamefont {Muhonen}, \citenamefont {Hudson}, \citenamefont {Itoh},
  \citenamefont {Morello},\ and\ \citenamefont {Dzurak}}]{veldhorst2014}%
  \BibitemOpen
  \bibfield  {author} {\bibinfo {author} {\bibfnamefont {M.}~\bibnamefont
  {Veldhorst}}, \bibinfo {author} {\bibfnamefont {J.~C.~C.}\ \bibnamefont
  {Hwang}}, \bibinfo {author} {\bibfnamefont {C.~H.}\ \bibnamefont {Yang}},
  \bibinfo {author} {\bibfnamefont {A.~W.}\ \bibnamefont {Leenstra}}, \bibinfo
  {author} {\bibfnamefont {B.}~\bibnamefont {de~Ronde}}, \bibinfo {author}
  {\bibfnamefont {J.~P.}\ \bibnamefont {Dehollain}}, \bibinfo {author}
  {\bibfnamefont {J.~T.}\ \bibnamefont {Muhonen}}, \bibinfo {author}
  {\bibfnamefont {F.~E.}\ \bibnamefont {Hudson}}, \bibinfo {author}
  {\bibfnamefont {K.~M.}\ \bibnamefont {Itoh}}, \bibinfo {author}
  {\bibfnamefont {A.}~\bibnamefont {Morello}},\ and\ \bibinfo {author}
  {\bibfnamefont {A.~S.}\ \bibnamefont {Dzurak}},\ }\bibfield  {title}
  {\bibinfo {title} {An addressable quantum dot qubit with fault-tolerant
  control-fidelity},\ }\href {http://dx.doi.org/10.1038/nnano.2014.216}
  {\bibfield  {journal} {\bibinfo  {journal} {Nat. Nanotechnol.}\ }\textbf
  {\bibinfo {volume} {9}},\ \bibinfo {pages} {981 EP } (\bibinfo {year}
  {2014})}\BibitemShut {NoStop}%
\bibitem [{\citenamefont {Hayes}\ \emph {et~al.}(2009)\citenamefont {Hayes},
  \citenamefont {Kiselev}, \citenamefont {Borselli}, \citenamefont {Bui},
  \citenamefont {III}, \citenamefont {Deelman}, \citenamefont {Maune},
  \citenamefont {Milosavljevic}, \citenamefont {Moon}, \citenamefont {Ross},
  \citenamefont {Schmitz}, \citenamefont {Gyure},\ and\ \citenamefont
  {Hunter}}]{hayes2009}%
  \BibitemOpen
  \bibfield  {author} {\bibinfo {author} {\bibfnamefont {R.~R.}\ \bibnamefont
  {Hayes}}, \bibinfo {author} {\bibfnamefont {A.~A.}\ \bibnamefont {Kiselev}},
  \bibinfo {author} {\bibfnamefont {M.~G.}\ \bibnamefont {Borselli}}, \bibinfo
  {author} {\bibfnamefont {S.~S.}\ \bibnamefont {Bui}}, \bibinfo {author}
  {\bibfnamefont {E.~T.~C.}\ \bibnamefont {III}}, \bibinfo {author}
  {\bibfnamefont {P.~W.}\ \bibnamefont {Deelman}}, \bibinfo {author}
  {\bibfnamefont {B.~M.}\ \bibnamefont {Maune}}, \bibinfo {author}
  {\bibfnamefont {I.}~\bibnamefont {Milosavljevic}}, \bibinfo {author}
  {\bibfnamefont {J.-S.}\ \bibnamefont {Moon}}, \bibinfo {author}
  {\bibfnamefont {R.~S.}\ \bibnamefont {Ross}}, \bibinfo {author}
  {\bibfnamefont {A.~E.}\ \bibnamefont {Schmitz}}, \bibinfo {author}
  {\bibfnamefont {M.~F.}\ \bibnamefont {Gyure}},\ and\ \bibinfo {author}
  {\bibfnamefont {A.~T.}\ \bibnamefont {Hunter}},\ }\bibfield  {title}
  {\bibinfo {title} {Lifetime measurements {($T_1$)} of electron spins in
  {Si/SiGe} quantum dots},\ }\href@noop {} {\  (\bibinfo {year} {2009})},\
  \Eprint {https://arxiv.org/abs/0908.0173} {arXiv:0908.0173
  [cond-mat.mes-hall]} \BibitemShut {NoStop}%
\bibitem [{\citenamefont {Prance}\ \emph {et~al.}(2012)\citenamefont {Prance},
  \citenamefont {Shi}, \citenamefont {Simmons}, \citenamefont {Savage},
  \citenamefont {Lagally}, \citenamefont {Schreiber}, \citenamefont
  {Vandersypen}, \citenamefont {Friesen}, \citenamefont {Joynt}, \citenamefont
  {Coppersmith} \emph {et~al.}}]{Prance2012}%
  \BibitemOpen
  \bibfield  {author} {\bibinfo {author} {\bibfnamefont {J.}~\bibnamefont
  {Prance}}, \bibinfo {author} {\bibfnamefont {Z.}~\bibnamefont {Shi}},
  \bibinfo {author} {\bibfnamefont {C.}~\bibnamefont {Simmons}}, \bibinfo
  {author} {\bibfnamefont {D.}~\bibnamefont {Savage}}, \bibinfo {author}
  {\bibfnamefont {M.}~\bibnamefont {Lagally}}, \bibinfo {author} {\bibfnamefont
  {L.}~\bibnamefont {Schreiber}}, \bibinfo {author} {\bibfnamefont
  {L.}~\bibnamefont {Vandersypen}}, \bibinfo {author} {\bibfnamefont
  {M.}~\bibnamefont {Friesen}}, \bibinfo {author} {\bibfnamefont
  {R.}~\bibnamefont {Joynt}}, \bibinfo {author} {\bibfnamefont
  {S.}~\bibnamefont {Coppersmith}}, \emph {et~al.},\ }\bibfield  {title}
  {\bibinfo {title} {Single-shot measurement of triplet-singlet relaxation in a
  {Si/SiGe} double quantum dot},\ }\href
  {https://doi.org/10.1103/PhysRevLett.108.046808} {\bibfield  {journal}
  {\bibinfo  {journal} {Phys. Rev. Lett.}\ }\textbf {\bibinfo {volume} {108}},\
  \bibinfo {pages} {046808} (\bibinfo {year} {2012})}\BibitemShut {NoStop}%
\bibitem [{\citenamefont {Chen}\ \emph {et~al.}(2021)\citenamefont {Chen},
  \citenamefont {Raach}, \citenamefont {Pan}, \citenamefont {Kiselev},
  \citenamefont {Acuna}, \citenamefont {Blumoff}, \citenamefont {Brecht},
  \citenamefont {Choi}, \citenamefont {Ha}, \citenamefont {Hulbert},
  \citenamefont {Jura}, \citenamefont {Keating}, \citenamefont {Noah},
  \citenamefont {Sun}, \citenamefont {Thomas}, \citenamefont {Borselli},
  \citenamefont {Jackson}, \citenamefont {Rakher},\ and\ \citenamefont
  {Ross}}]{chen2021}%
  \BibitemOpen
  \bibfield  {author} {\bibinfo {author} {\bibfnamefont {E.~H.}\ \bibnamefont
  {Chen}}, \bibinfo {author} {\bibfnamefont {K.}~\bibnamefont {Raach}},
  \bibinfo {author} {\bibfnamefont {A.}~\bibnamefont {Pan}}, \bibinfo {author}
  {\bibfnamefont {A.~A.}\ \bibnamefont {Kiselev}}, \bibinfo {author}
  {\bibfnamefont {E.}~\bibnamefont {Acuna}}, \bibinfo {author} {\bibfnamefont
  {J.~Z.}\ \bibnamefont {Blumoff}}, \bibinfo {author} {\bibfnamefont
  {T.}~\bibnamefont {Brecht}}, \bibinfo {author} {\bibfnamefont {M.~D.}\
  \bibnamefont {Choi}}, \bibinfo {author} {\bibfnamefont {W.}~\bibnamefont
  {Ha}}, \bibinfo {author} {\bibfnamefont {D.~R.}\ \bibnamefont {Hulbert}},
  \bibinfo {author} {\bibfnamefont {M.~P.}\ \bibnamefont {Jura}}, \bibinfo
  {author} {\bibfnamefont {T.~E.}\ \bibnamefont {Keating}}, \bibinfo {author}
  {\bibfnamefont {R.}~\bibnamefont {Noah}}, \bibinfo {author} {\bibfnamefont
  {B.}~\bibnamefont {Sun}}, \bibinfo {author} {\bibfnamefont {B.~J.}\
  \bibnamefont {Thomas}}, \bibinfo {author} {\bibfnamefont {M.~G.}\
  \bibnamefont {Borselli}}, \bibinfo {author} {\bibfnamefont {C.}~\bibnamefont
  {Jackson}}, \bibinfo {author} {\bibfnamefont {M.~T.}\ \bibnamefont
  {Rakher}},\ and\ \bibinfo {author} {\bibfnamefont {R.~S.}\ \bibnamefont
  {Ross}},\ }\bibfield  {title} {\bibinfo {title} {Detuning axis pulsed
  spectroscopy of valley-orbital states in
  $\mathrm{Si}$/$\mathrm{Si}$-$\mathrm{Ge}$ quantum dots},\ }\href
  {https://doi.org/10.1103/PhysRevApplied.15.044033} {\bibfield  {journal}
  {\bibinfo  {journal} {Phys. Rev. Appl.}\ }\textbf {\bibinfo {volume} {15}},\
  \bibinfo {pages} {044033} (\bibinfo {year} {2021})}\BibitemShut {NoStop}%
\bibitem [{\citenamefont {Melnikov}\ and\ \citenamefont
  {Leburton}(2006)}]{melnikov2006}%
  \BibitemOpen
  \bibfield  {author} {\bibinfo {author} {\bibfnamefont {D.~V.}\ \bibnamefont
  {Melnikov}}\ and\ \bibinfo {author} {\bibfnamefont {J.-P.}\ \bibnamefont
  {Leburton}},\ }\bibfield  {title} {\bibinfo {title} {Dimensionality effects
  in the two-electron system in circular and elliptic quantum dots},\ }\href
  {https://doi.org/10.1103/PhysRevB.73.085320} {\bibfield  {journal} {\bibinfo
  {journal} {Phys. Rev. B}\ }\textbf {\bibinfo {volume} {73}},\ \bibinfo
  {pages} {085320} (\bibinfo {year} {2006})}\BibitemShut {NoStop}%
\bibitem [{\citenamefont {Weinmann}\ \emph {et~al.}(1994)\citenamefont
  {Weinmann}, \citenamefont {Häusler}, \citenamefont {Pfaff}, \citenamefont
  {Kramer},\ and\ \citenamefont {Weiss}}]{weinmann1994}%
  \BibitemOpen
  \bibfield  {author} {\bibinfo {author} {\bibfnamefont {D.}~\bibnamefont
  {Weinmann}}, \bibinfo {author} {\bibfnamefont {W.}~\bibnamefont {Häusler}},
  \bibinfo {author} {\bibfnamefont {W.}~\bibnamefont {Pfaff}}, \bibinfo
  {author} {\bibfnamefont {B.}~\bibnamefont {Kramer}},\ and\ \bibinfo {author}
  {\bibfnamefont {U.}~\bibnamefont {Weiss}},\ }\bibfield  {title} {\bibinfo
  {title} {Spin blockade in non-linear transport through quantum dots},\ }\href
  {https://doi.org/10.1209/0295-5075/26/6/012} {\bibfield  {journal} {\bibinfo
  {journal} {EPL}\ }\textbf {\bibinfo {volume} {26}},\ \bibinfo {pages} {467}
  (\bibinfo {year} {1994})}\BibitemShut {NoStop}%
\bibitem [{\citenamefont {Studenikin}\ \emph {et~al.}(2012)\citenamefont
  {Studenikin}, \citenamefont {Thorgrimson}, \citenamefont {Aers},
  \citenamefont {Kam}, \citenamefont {Zawadzki}, \citenamefont {Wasilewski},
  \citenamefont {Bogan},\ and\ \citenamefont {Sachrajda}}]{studenikin2012}%
  \BibitemOpen
  \bibfield  {author} {\bibinfo {author} {\bibfnamefont {S.}~\bibnamefont
  {Studenikin}}, \bibinfo {author} {\bibfnamefont {J.}~\bibnamefont
  {Thorgrimson}}, \bibinfo {author} {\bibfnamefont {G.}~\bibnamefont {Aers}},
  \bibinfo {author} {\bibfnamefont {A.}~\bibnamefont {Kam}}, \bibinfo {author}
  {\bibfnamefont {P.}~\bibnamefont {Zawadzki}}, \bibinfo {author}
  {\bibfnamefont {Z.}~\bibnamefont {Wasilewski}}, \bibinfo {author}
  {\bibfnamefont {A.}~\bibnamefont {Bogan}},\ and\ \bibinfo {author}
  {\bibfnamefont {A.}~\bibnamefont {Sachrajda}},\ }\bibfield  {title} {\bibinfo
  {title} {Enhanced charge detection of spin qubit readout via an intermediate
  state},\ }\href {https://doi.org/10.1063/1.4749281} {\bibfield  {journal}
  {\bibinfo  {journal} {Appl. Phys. Lett.}\ }\textbf {\bibinfo {volume}
  {101}},\ \bibinfo {pages} {233101} (\bibinfo {year} {2012})}\BibitemShut
  {NoStop}%
\bibitem [{\citenamefont {Mason}\ \emph {et~al.}(2015)\citenamefont {Mason},
  \citenamefont {Studenikin}, \citenamefont {Kam}, \citenamefont {Wasilewski},
  \citenamefont {Sachrajda},\ and\ \citenamefont {Kycia}}]{mason2015}%
  \BibitemOpen
  \bibfield  {author} {\bibinfo {author} {\bibfnamefont {J.}~\bibnamefont
  {Mason}}, \bibinfo {author} {\bibfnamefont {S.}~\bibnamefont {Studenikin}},
  \bibinfo {author} {\bibfnamefont {A.}~\bibnamefont {Kam}}, \bibinfo {author}
  {\bibfnamefont {Z.}~\bibnamefont {Wasilewski}}, \bibinfo {author}
  {\bibfnamefont {A.}~\bibnamefont {Sachrajda}},\ and\ \bibinfo {author}
  {\bibfnamefont {J.}~\bibnamefont {Kycia}},\ }\bibfield  {title} {\bibinfo
  {title} {Role of metastable charge states in a quantum-dot spin-qubit
  readout},\ }\href {https://doi.org/10.1103/PhysRevB.92.125434} {\bibfield
  {journal} {\bibinfo  {journal} {Phys. Rev. B}\ }\textbf {\bibinfo {volume}
  {92}},\ \bibinfo {pages} {125434} (\bibinfo {year} {2015})}\BibitemShut
  {NoStop}%
\bibitem [{\citenamefont {Harvey-Collard}\ \emph {et~al.}(2018)\citenamefont
  {Harvey-Collard}, \citenamefont {D'Anjou}, \citenamefont {Rudolph},
  \citenamefont {Jacobson}, \citenamefont {Dominguez}, \citenamefont
  {Ten~Eyck}, \citenamefont {Wendt}, \citenamefont {Pluym}, \citenamefont
  {Lilly}, \citenamefont {Coish}, \citenamefont {Pioro-Ladri\`ere},\ and\
  \citenamefont {Carroll}}]{harveycollard2018}%
  \BibitemOpen
  \bibfield  {author} {\bibinfo {author} {\bibfnamefont {P.}~\bibnamefont
  {Harvey-Collard}}, \bibinfo {author} {\bibfnamefont {B.}~\bibnamefont
  {D'Anjou}}, \bibinfo {author} {\bibfnamefont {M.}~\bibnamefont {Rudolph}},
  \bibinfo {author} {\bibfnamefont {N.~T.}\ \bibnamefont {Jacobson}}, \bibinfo
  {author} {\bibfnamefont {J.}~\bibnamefont {Dominguez}}, \bibinfo {author}
  {\bibfnamefont {G.~A.}\ \bibnamefont {Ten~Eyck}}, \bibinfo {author}
  {\bibfnamefont {J.~R.}\ \bibnamefont {Wendt}}, \bibinfo {author}
  {\bibfnamefont {T.}~\bibnamefont {Pluym}}, \bibinfo {author} {\bibfnamefont
  {M.~P.}\ \bibnamefont {Lilly}}, \bibinfo {author} {\bibfnamefont {W.~A.}\
  \bibnamefont {Coish}}, \bibinfo {author} {\bibfnamefont {M.}~\bibnamefont
  {Pioro-Ladri\`ere}},\ and\ \bibinfo {author} {\bibfnamefont {M.~S.}\
  \bibnamefont {Carroll}},\ }\bibfield  {title} {\bibinfo {title}
  {High-fidelity single-shot readout for a spin qubit via an enhanced latching
  mechanism},\ }\href {https://doi.org/10.1103/PhysRevX.8.021046} {\bibfield
  {journal} {\bibinfo  {journal} {Phys. Rev. X}\ }\textbf {\bibinfo {volume}
  {8}},\ \bibinfo {pages} {021046} (\bibinfo {year} {2018})}\BibitemShut
  {NoStop}%
\bibitem [{\citenamefont {van Diepen}\ \emph {et~al.}(2021)\citenamefont {van
  Diepen}, \citenamefont {Hsiao}, \citenamefont {Mukhopadhyay}, \citenamefont
  {Reichl}, \citenamefont {Wegscheider},\ and\ \citenamefont
  {Vandersypen}}]{van2020}%
  \BibitemOpen
  \bibfield  {author} {\bibinfo {author} {\bibfnamefont {C.~J.}\ \bibnamefont
  {van Diepen}}, \bibinfo {author} {\bibfnamefont {T.-K.}\ \bibnamefont
  {Hsiao}}, \bibinfo {author} {\bibfnamefont {U.}~\bibnamefont {Mukhopadhyay}},
  \bibinfo {author} {\bibfnamefont {C.}~\bibnamefont {Reichl}}, \bibinfo
  {author} {\bibfnamefont {W.}~\bibnamefont {Wegscheider}},\ and\ \bibinfo
  {author} {\bibfnamefont {L.~M.}\ \bibnamefont {Vandersypen}},\ }\bibfield
  {title} {\bibinfo {title} {Electron cascade for distant spin readout},\
  }\href {https://doi.org/10.1038/s41467-020-20388-6} {\bibfield  {journal}
  {\bibinfo  {journal} {Nat. Commun.}\ }\textbf {\bibinfo {volume} {12}},\
  \bibinfo {pages} {1} (\bibinfo {year} {2021})}\BibitemShut {NoStop}%
\bibitem [{\citenamefont {Beenakker}(1991)}]{beenakker1991}%
  \BibitemOpen
  \bibfield  {author} {\bibinfo {author} {\bibfnamefont {C.~W.~J.}\
  \bibnamefont {Beenakker}},\ }\bibfield  {title} {\bibinfo {title} {Theory of
  {Coulomb}-blockade oscillations in the conductance of a quantum dot},\ }\href
  {https://doi.org/10.1103/PhysRevB.44.1646} {\bibfield  {journal} {\bibinfo
  {journal} {Phys. Rev. B}\ }\textbf {\bibinfo {volume} {44}},\ \bibinfo
  {pages} {1646} (\bibinfo {year} {1991})}\BibitemShut {NoStop}%
\bibitem [{\citenamefont {Van~Houten}\ \emph {et~al.}(1992)\citenamefont
  {Van~Houten}, \citenamefont {Beenakker},\ and\ \citenamefont
  {Staring}}]{van1992}%
  \BibitemOpen
  \bibfield  {author} {\bibinfo {author} {\bibfnamefont {H.}~\bibnamefont
  {Van~Houten}}, \bibinfo {author} {\bibfnamefont {C.}~\bibnamefont
  {Beenakker}},\ and\ \bibinfo {author} {\bibfnamefont {A.}~\bibnamefont
  {Staring}},\ }\bibfield  {title} {\bibinfo {title} {{C}oulomb-blockade
  oscillations in semiconductor nanostructures},\ }in\ \href
  {https://link.springer.com/chapter/10.1007/978-1-4757-2166-9_5} {\emph
  {\bibinfo {booktitle} {Single charge tunneling}}}\ (\bibinfo  {publisher}
  {Springer},\ \bibinfo {year} {1992})\ pp.\ \bibinfo {pages}
  {167--216}\BibitemShut {NoStop}%
\bibitem [{\citenamefont {Horibe}\ \emph {et~al.}(2015)\citenamefont {Horibe},
  \citenamefont {Kodera},\ and\ \citenamefont {Oda}}]{horibe2015}%
  \BibitemOpen
  \bibfield  {author} {\bibinfo {author} {\bibfnamefont {K.}~\bibnamefont
  {Horibe}}, \bibinfo {author} {\bibfnamefont {T.}~\bibnamefont {Kodera}},\
  and\ \bibinfo {author} {\bibfnamefont {S.}~\bibnamefont {Oda}},\ }\bibfield
  {title} {\bibinfo {title} {Back-action-induced excitation of electrons in a
  silicon quantum dot with a single-electron transistor charge sensor},\ }\href
  {https://doi.org/10.1063/1.4907894} {\bibfield  {journal} {\bibinfo
  {journal} {Appl. Phys. Lett.}\ }\textbf {\bibinfo {volume} {106}},\ \bibinfo
  {pages} {053119} (\bibinfo {year} {2015})}\BibitemShut {NoStop}%
\bibitem [{\citenamefont {Li}\ \emph {et~al.}(2012)\citenamefont {Li},
  \citenamefont {Xiao}, \citenamefont {Cao}, \citenamefont {Zhou},
  \citenamefont {Shang}, \citenamefont {Tu}, \citenamefont {Guo}, \citenamefont
  {Jiang},\ and\ \citenamefont {Guo}}]{li2012}%
  \BibitemOpen
  \bibfield  {author} {\bibinfo {author} {\bibfnamefont {H.}~\bibnamefont
  {Li}}, \bibinfo {author} {\bibfnamefont {M.}~\bibnamefont {Xiao}}, \bibinfo
  {author} {\bibfnamefont {G.}~\bibnamefont {Cao}}, \bibinfo {author}
  {\bibfnamefont {C.}~\bibnamefont {Zhou}}, \bibinfo {author} {\bibfnamefont
  {R.}~\bibnamefont {Shang}}, \bibinfo {author} {\bibfnamefont
  {T.}~\bibnamefont {Tu}}, \bibinfo {author} {\bibfnamefont {G.}~\bibnamefont
  {Guo}}, \bibinfo {author} {\bibfnamefont {H.}~\bibnamefont {Jiang}},\ and\
  \bibinfo {author} {\bibfnamefont {G.}~\bibnamefont {Guo}},\ }\bibfield
  {title} {\bibinfo {title} {Back-action-induced non-equilibrium effect in
  electron charge counting statistics},\ }\href
  {https://doi.org/10.1063/1.3691255} {\bibfield  {journal} {\bibinfo
  {journal} {Appl. Phys. Lett.}\ }\textbf {\bibinfo {volume} {100}},\ \bibinfo
  {pages} {092112} (\bibinfo {year} {2012})}\BibitemShut {NoStop}%
\bibitem [{\citenamefont {Culcer}\ \emph {et~al.}(2010)\citenamefont {Culcer},
  \citenamefont {Cywi\ifmmode~\acute{n}\else \'{n}\fi{}ski}, \citenamefont
  {Li}, \citenamefont {Hu},\ and\ \citenamefont {Das~Sarma}}]{dassarma2010}%
  \BibitemOpen
  \bibfield  {author} {\bibinfo {author} {\bibfnamefont {D.}~\bibnamefont
  {Culcer}}, \bibinfo {author} {\bibfnamefont {L.}~\bibnamefont
  {Cywi\ifmmode~\acute{n}\else \'{n}\fi{}ski}}, \bibinfo {author}
  {\bibfnamefont {Q.}~\bibnamefont {Li}}, \bibinfo {author} {\bibfnamefont
  {X.}~\bibnamefont {Hu}},\ and\ \bibinfo {author} {\bibfnamefont
  {S.}~\bibnamefont {Das~Sarma}},\ }\bibfield  {title} {\bibinfo {title}
  {Quantum dot spin qubits in silicon: multivalley physics},\ }\href
  {https://doi.org/10.1103/PhysRevB.82.155312} {\bibfield  {journal} {\bibinfo
  {journal} {Phys. Rev. B}\ }\textbf {\bibinfo {volume} {82}},\ \bibinfo
  {pages} {155312} (\bibinfo {year} {2010})}\BibitemShut {NoStop}%
\bibitem [{\citenamefont {Stepanenko}\ \emph {et~al.}(2012)\citenamefont
  {Stepanenko}, \citenamefont {Rudner}, \citenamefont {Halperin},\ and\
  \citenamefont {Loss}}]{stepanenko2012}%
  \BibitemOpen
  \bibfield  {author} {\bibinfo {author} {\bibfnamefont {D.}~\bibnamefont
  {Stepanenko}}, \bibinfo {author} {\bibfnamefont {M.}~\bibnamefont {Rudner}},
  \bibinfo {author} {\bibfnamefont {B.~I.}\ \bibnamefont {Halperin}},\ and\
  \bibinfo {author} {\bibfnamefont {D.}~\bibnamefont {Loss}},\ }\bibfield
  {title} {\bibinfo {title} {Singlet-triplet splitting in double quantum dots
  due to spin-orbit and hyperfine interactions},\ }\href
  {https://doi.org/10.1103/PhysRevB.85.075416} {\bibfield  {journal} {\bibinfo
  {journal} {Phys. Rev. B}\ }\textbf {\bibinfo {volume} {85}},\ \bibinfo
  {pages} {075416} (\bibinfo {year} {2012})}\BibitemShut {NoStop}%
\bibitem [{\citenamefont {Whittaker}\ \emph {et~al.}(2014)\citenamefont
  {Whittaker}, \citenamefont {Da~Silva}, \citenamefont {Allman}, \citenamefont
  {Lecocq}, \citenamefont {Cicak}, \citenamefont {Sirois}, \citenamefont
  {Teufel}, \citenamefont {Aumentado},\ and\ \citenamefont
  {Simmonds}}]{whittaker2014}%
  \BibitemOpen
  \bibfield  {author} {\bibinfo {author} {\bibfnamefont {J.~D.}\ \bibnamefont
  {Whittaker}}, \bibinfo {author} {\bibfnamefont {F.}~\bibnamefont {Da~Silva}},
  \bibinfo {author} {\bibfnamefont {M.~S.}\ \bibnamefont {Allman}}, \bibinfo
  {author} {\bibfnamefont {F.}~\bibnamefont {Lecocq}}, \bibinfo {author}
  {\bibfnamefont {K.}~\bibnamefont {Cicak}}, \bibinfo {author} {\bibfnamefont
  {A.}~\bibnamefont {Sirois}}, \bibinfo {author} {\bibfnamefont
  {J.}~\bibnamefont {Teufel}}, \bibinfo {author} {\bibfnamefont
  {J.}~\bibnamefont {Aumentado}},\ and\ \bibinfo {author} {\bibfnamefont
  {R.~W.}\ \bibnamefont {Simmonds}},\ }\bibfield  {title} {\bibinfo {title}
  {Tunable-cavity {QED} with phase qubits},\ }\href
  {https://doi.org/10.1103/PhysRevB.90.024513} {\bibfield  {journal} {\bibinfo
  {journal} {Phys. Rev.d B}\ }\textbf {\bibinfo {volume} {90}},\ \bibinfo
  {pages} {024513} (\bibinfo {year} {2014})}\BibitemShut {NoStop}%
\bibitem [{\citenamefont {Myerson}\ \emph {et~al.}(2008)\citenamefont
  {Myerson}, \citenamefont {Szwer}, \citenamefont {Webster}, \citenamefont
  {Allcock}, \citenamefont {Curtis}, \citenamefont {Imreh}, \citenamefont
  {Sherman}, \citenamefont {Stacey}, \citenamefont {Steane},\ and\
  \citenamefont {Lucas}}]{myerson2008}%
  \BibitemOpen
  \bibfield  {author} {\bibinfo {author} {\bibfnamefont {A.}~\bibnamefont
  {Myerson}}, \bibinfo {author} {\bibfnamefont {D.}~\bibnamefont {Szwer}},
  \bibinfo {author} {\bibfnamefont {S.}~\bibnamefont {Webster}}, \bibinfo
  {author} {\bibfnamefont {D.}~\bibnamefont {Allcock}}, \bibinfo {author}
  {\bibfnamefont {M.}~\bibnamefont {Curtis}}, \bibinfo {author} {\bibfnamefont
  {G.}~\bibnamefont {Imreh}}, \bibinfo {author} {\bibfnamefont
  {J.}~\bibnamefont {Sherman}}, \bibinfo {author} {\bibfnamefont
  {D.}~\bibnamefont {Stacey}}, \bibinfo {author} {\bibfnamefont
  {A.}~\bibnamefont {Steane}},\ and\ \bibinfo {author} {\bibfnamefont
  {D.}~\bibnamefont {Lucas}},\ }\bibfield  {title} {\bibinfo {title}
  {High-fidelity readout of trapped-ion qubits},\ }\href
  {https://doi.org/10.1103/PhysRevLett.100.200502} {\bibfield  {journal}
  {\bibinfo  {journal} {Phys. Rev. Lett.}\ }\textbf {\bibinfo {volume} {100}},\
  \bibinfo {pages} {200502} (\bibinfo {year} {2008})}\BibitemShut {NoStop}%
\end{thebibliography}%
\cleardoublepage
\newpage
\appendix
\section{High-speed Adder Circuit}\label{section:app-hsa}

\begin{figure}
	\includegraphics[width=1\linewidth]{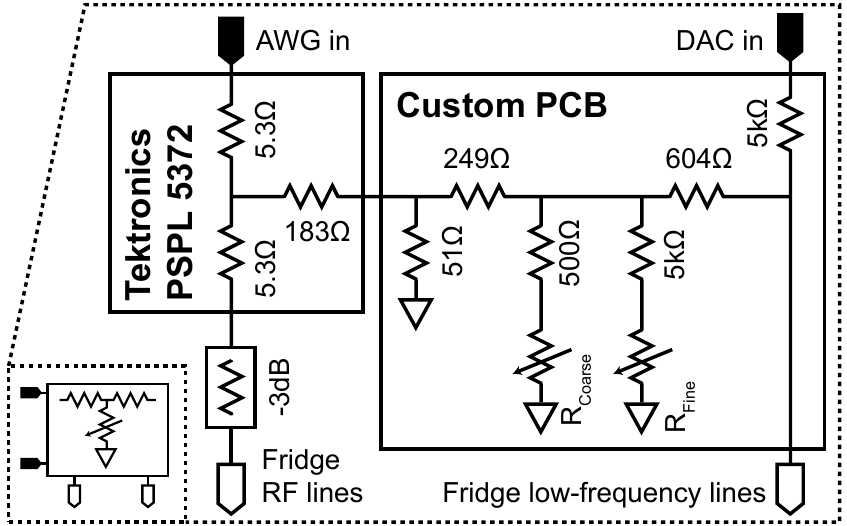} 
	\caption{
		\textbf{Room temperature high speed adder circuit.} 
		This circuit allows high-bandwidth but baseband control, combining inputs both from NI AWGs and custom low-noise DACs.   
		Together with the fridge components shown in \reffig{fig:signal_chain} (attenuation, bias tee and low-pass filter), this forms the full signal chain,  providing the benefits discussed in the text.  
		$R_{\text{fine}}$ and $R_{\text{coarse}}$ are highly-stable potentiometers that allow for balancing the attenuation of the high-frequency and low-frequency signal paths.}
	\label{fig:hsa}
\end{figure}

Sending high-bandwidth baseband signals into a dilution refrigerator is a 
non-trivial problem.
To protect sensitive quantum devices from thermal Johnson noise and instrumentation noise, microwave or all-AC signals are generally attenuated heavily at intermediate temperature stages within a refrigerator.
However, it is not desirable to send large DC signals through attenuators, as they will constantly dissipate power inside of the cryostat.
Fortunately we can instead aggressively low-pass filter DC signals.
In our adder circuit we use both of these techniques---sending the AC and DC components of a signal in separately, which are then combined at a base-temperature bias tee (see \reffig{fig:hsa}).

An added complication is that the baseband pulses (generated from NI5451 AWGs) contain both AC and DC components.
We get around this by splitting the AWG signal. 
The AC component travels directly through the (attenuated) AC lines.
The DC component is attenuated at room temperature via a simple resistive divider, then combined with the true DC signal from a low-noise DAC, before traveling through the (filtered) DC lines.
It is critical that the attenuation of both of these paths is matched, or the mismatch will result in long-timescale waveform distortion.  
This is particularly evident in the shift of measurement signals during long experimental sequences, such as randomized benchmarking.
Treated improperly, this shift of signal with time can be mistaken as resulting from some other dot physics.
We test this using the stabilization of a dot loading line bias after a step-edge pulse, and the attenuation of the resistive divider is fine-tuned with a highly-stable Vishay potentiometer.

\section{Contributions to SNR}\label{section:app-SNR}
\subsection{Noise amplitude}\label{section:app-disc_noise}
The budget for measurement noise is dominated by three effectively white noise sources of comparable magnitude (input-referred HEMT noise, sense-resistor Johnson noise, DCS current shot noise) and by the noise on the DCS potential.
In our measurement chain configuration, the former three sources each generate a noise spectral density of roughly 200-300 pV/$\sqrt{\text{Hz}}$.
The quantum dot charge noise spectrum is typically of the form $A^2/f$, with effective gate-referred voltage spectral density $A \approx 5~\mu$V/$\sqrt{\text{Hz}}$ at 1 Hz.

The effect of this noise source on readout depends on how the DCS is operated.
In particular, two extreme choices are to bias the DCS gate voltage to maximize either the transconductance $G_m$ (peak of the derivative of conductance with respect to M voltage) or the conductance (the conventional Coulomb blockade peak point). 
The former maximizes the signal when the magnitude of the charge-dependent potential shift at the DCS (\refapp{section:app-disc_signal}) is small compared to the width of the Coulomb peak, and we can approximate the effect of the DCS-potential noise as a linear shift of the conductance.
The latter maximizes the signal when the magnitude of the charge-dependent potential shift is much larger than the width of the Coulomb peak and the state-dependent conductance difference is maximal.
In this case, we are insensitive to DCS potential noise to first order.
From these two extrema it is evident that the DCS bias that maximizes signal may not be the correct bias to maximize SNR, particularly in the $1/f$-dominated long-integration regime.

To predict the ultimate expected noise amplitude we integrate the white noise sources with the filter function appropriate for our square-wave modulated measurement tone, yielding a current noise of $\sigma_{SD}^2$.
We account for the charge noise in a conservative analysis, taking the small-shift approximation such that, to first order, the output measured current is modulated by $G_m d(t)$, where $d(t)$ is the (noisy) potential shift.
If we add that to the previous analysis, assume $d(t)$ has a power-spectral density given by $A^2/f$, and account for our averaging scheme when incorporating an (approximately instantaneously) subsequent reference measurement, the total histogram variance can be expressed as
\begin{equation} \label{eqn:noise_spectrum}
	\sigma_{\text{I}}^2=2\sigma_{SD}^2+16\ln 2 G_m^2 A^2.
\end{equation}
This sets an upper bound on the effects of DCS charge noise on measurement histograms.
For relevant measurement parameters, we predict $\sigma_I\approx$ 5pA at long duration, which is in line with observed values (e.g. \reffig {fig:snr_and_hist}b, though note in that figure the current values for referenced measurements have been divided by a factor of 4 for inconsequential implementation details.)

As subtracting a reference measurement increases the white noise contribution, it is only preferable when the noise is dominated by $1/f$ or other slow processes.  
This can occur either because the $1/f$ contribution is generally large or because one seeks the highest possible SNR and is willing to average long enough to reduce the white noise contribution.
Even in such a regime, performing a reference measurement for every experimental query is unlikely to be optimal.

\subsection{Signal amplitude}\label{section:app-disc_signal}
We can estimate the magnitude of the HEMT-input-referred voltage signal as a product of five factors:
\begin{enumerate}
	\item the spin-to-charge conversion efficiency,
	\item the inter-charge-state DCS potential shift,
	\item the DCS transconductance $G_m$,
	\item the maximum applicable source-drain bias, and
	\item the value of the sense resistor.
\end{enumerate}
The spin-to-charge conversion efficiency describes how effectively the spin states correspond to different charge states at the measurement bias, and approaches unity when the two-electron excited-state splitting is larger than the singlet- or triplet-selective charge transition widths.
That width can be reduced by decreasing the effective electron temperature and tunnel coupling, but it is also sensitive to charge noise (see the supplemental material of \cite{jones2019}).
In practice, we find an excited-state splitting of roughly 100~$\mu$eV is sufficient to push this term near unity at typical operating electron temperature and tunnel coupling.
The magnitude of the potential shift is a straightforward calculation from the device geometry and Coulomb's law, accounting for the effects of image charges due to the electrostatic environment.
The transconductance around a Coulomb peak is a well-studied question, depending on tunneling rates of the DCS tunnel barriers and the effective electron temperature \cite{beenakker1991,van1992}.

The maximum applicable source bias is limited by two factors.
First (as noted in \refsec{section:results_snr}) nonlinearity in the DCS causes the conductance to decrease after a critical value.
Second, we also observe that at sufficiently large source-drain bias, the spin relaxation lifetime begins to decrease, discussed in \refapp{section:app-t1sd}.
The voltage signal can be linearly increased with high-resistance sense resistors, but at the expense of increased Johnson and shot noise.
Additionally, due to parasitic capacitance, the accessible measurement frequency decreases quickly, limiting the ability to mitigate low-frequency noise.

\section{$T_1$ vs Source-Drain Bias Amplitude}\label{section:app-t1sd}
\begin{center}
	\begin{figure}[]
		\centering
		\includegraphics[width=1\linewidth]{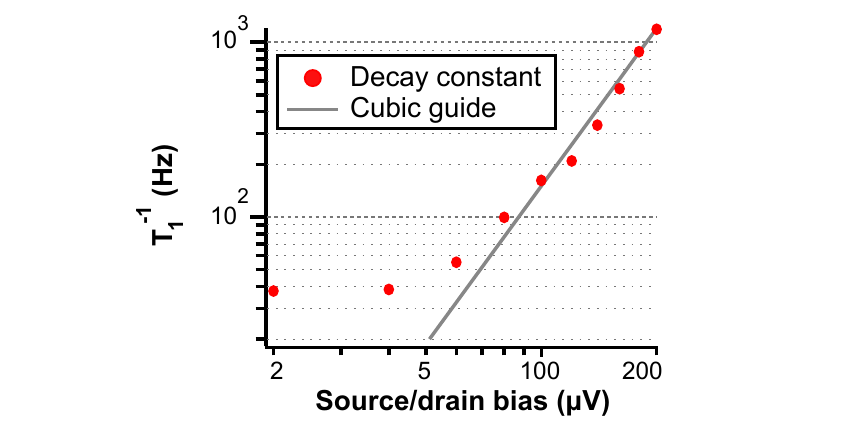} 
		\caption{
			\textbf{$T_1$ as a function of applied measurement bias}.   
			The decay rate increases after a certain bias amplitude of energy comparable to 
			the singlet-triplet splitting.  
			We also provide a cubic guide to the eye, which appears to track the asymptotic behavior.
		}
		\label{fig:t1_vs_amplitude}
	\end{figure}
\end{center}

We have also explored the effect of measurement-bias-driven $T_1$ decay using the experiment discussed in \refsec{section:disc_t1}.  
\reffig{fig:t1_vs_amplitude} shows that $T_1$ decreases strongly when we bias beyond roughly 75~$\mu$V, a value suggestively close in correspondence to the S-T splitting at the middle of the measurement window, e.g., equal to the gap at that detuning between the ground (2,0) state and the lowest (1,1) excited triplet state. 
These observations show similarities to other reports in the literature \cite{horibe2015,li2012}. 
The onset value and the nonlinear scaling with bias are potentially indicative of enhanced relaxation due to phonons generated by the enhanced dissipation of the DCS. 
Further work is needed to understand the microscopic underpinnings of this effect.

\section{Measure Window Energy Spectrum}\label{section:app-measwindow}
The level structure in the vicinity of a (1,1)-(2,0) anti-crossing is quite complex, especially in Si/SiGe quantum dots where the valley degree of freedom introduces additional excitations. 
The valley splitting $E_v$ is usually 10s-100s~$\mu$eV while the single-electron orbital energies tend to be 1 meV and higher. 
However, the two-electron (2e) orbital energy $E_o$ can be considerably smaller (on order of 10s-100s~$\mu$eV) than the valley splitting due to the interplay of the Coulomb interaction and spatial asymmetry. 
A simplified model spectrum for the device in this paper is depicted in \reffig{fig:spectrum_and_sbs}a. 
The first excited two electron orbital state in a single dot is a spin triplet and hence three-fold degenerate in the absence of a magnetic field. 
By contrast, a two electron valley excitation can be either singlet or triplet and is nearly fourfold degenerate because of the negligible intervalley exchange interaction~\cite{dassarma2010}. 
These degeneracies alter the occupation of excited states and hence the initialization fidelity, as reflected in the partition function of  \refeq{eqn:partition_function}. 
The level structure in practice will be further complicated by the presence of other doubly-excited states, valley-spin-orbit mixing, and Zeeman splitting of polarized triplets due to external magnetic fields.

This energy spectrum also provides some hints about the physics of $T_1$ relaxation within the measurement window, which requires both spin- and charge-mixing of the singlet and triplet states. 
Thus, for example, the hyperfine interaction can couple the T$_{x}$(1,1) states to the nearly degenerate and parallel excited S(1,1) state, which can then decay via phonon or photon emission to the ground S(2,0) state. 
The spin-orbit interaction likewise introduces magnetic gradients mixing T(1,1) and S(1,1) as well as spin-flip tunneling, which couples T$_x$(1,1) to S(2,0) directly, both of which are sensitive to the magnetic field orientation~\cite{stepanenko2012}. 
Similarly the reverse processes are also possible.
These terms will create additional excited singlet-triplet anticrossings within the measure window when the polarized triplet states are split by external magnetic fields (not pictured), which we believe are connected to the $T_1$ ``hot spots'' visible in \reffig{fig:t1_snr_vs_detuning}. 
These features are qualitatively reproduced in calculations directly diagonalizing the (1,1)-(2,0) Hamiltonian including spin-mixing terms, though quantitative comparison is currently inhibited because several key parameters (particularly for the spin-orbit coupling) have not been characterized.
In such a model, the relaxation rate may be sensitive in both magnitude and hot spot location to magnetic field, tunnel coupling, and valley mixing, consistent with the variable, tuneup- and device-dependent $T_1$ features we observe experimentally.

\section{Exchange oscillation contrast}\label{section:app-exchange}
\begin{figure}[h]
	\includegraphics[width=1\linewidth]{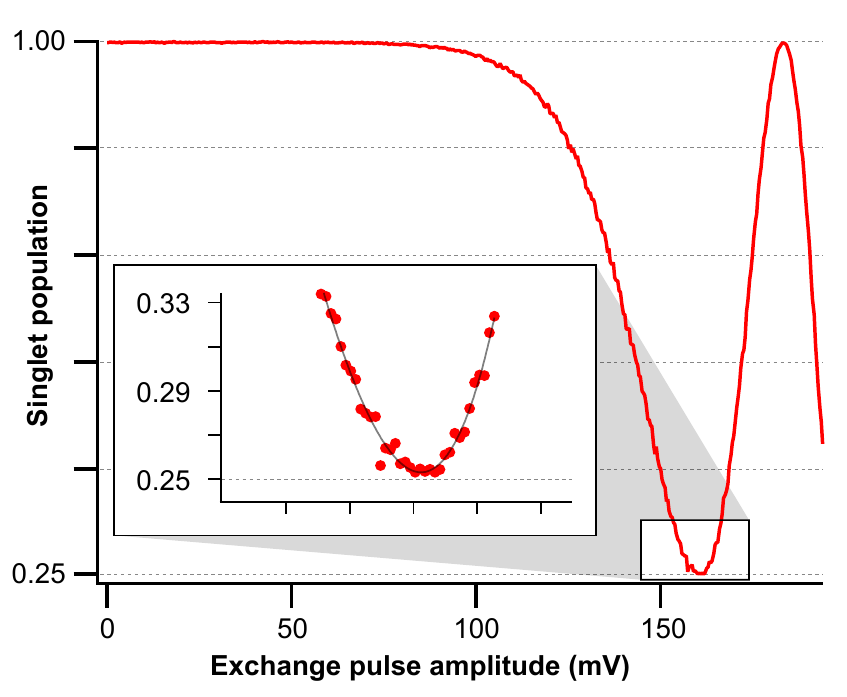} 
	\caption{
		\textbf{Exchange oscillation contrast.} 
		Exchange oscillations on the $n$ axis, plotting thresholded singlet population vs magnitude of throw on the symmetric axis, referenced to the voltage change on the X2 gate.  
		The $n$ axis should rotate the logical 0 state from 100\% singlet to 25\% singlet.  
		The inset has magnified the same data near first near-25\% trough.  
		Interpreted directly, half of the difference between this contrast and 0.75 would indicate a SPAM infidelity of $2.8\expn{-3}$.
	}
	\label{fig:exchange-oscillations}
\end{figure}

The contrast of exchange oscillations is another plausible joint SPAM performance metric in our system.
We prefer $\fspam$ from benchmarking for the reasons noted in the text, but in this case the resulting estimates are quite similar (see \reffig{fig:exchange-oscillations}).
In other physical architectures this approach might be compared to the contrast of Rabi oscillations.

\section{Definition of $\fspam$}\label{section:app-fspam}
This integrated preparation and measurement metric  (discussed in \refsec{section:joint}) was first presented in Ref.~\onlinecite{andrews2019}).
$\fspam$ is derived from ``blind'' randomized benchmarking, a procedure in which an ordinary randomized benchmarking experiment (with sequences that compile to the identity) is executed along with a second benchmarking experiment, whose sequences are instead engineered to compile to a population-inverting gate.
We then fit the results of these two experiments to the form
\begin{subequations}
	\begin{align}
		y_{\text{0}}&=A+B\left(1-p\right)^N+C\left(1-q\right)^N \label{eqn:brb0} \\
		y_{\text{1}}&=A-B\left(1-p\right)^N+C\left(1-q\right)^N \label{eqn:brb1}.
	\end{align}
\end{subequations}
To assess SPAM performance, we evaluate the contrast of these fit curves at zero Clifford operations, which attempts to subtract any contrast loss due to imperfect qubit rotations.
The SPAM infidelity is given by half of the missing contrast,
\begin{equation}
	1-\fspam := 0.5-B \label{eqn:brbfid}.
\end{equation}

\section{Mapping errors}\label{section:app-mapping}
During DFS operation, bias configurations used for initialization and measurement are different than the idle bias used in-between coherent manipulations. 
Transitioning between these operating points can give rise to ``mapping'' errors.
Similar mapping steps are not uncommon in other architectures, e.g., flux-tunable superconducting qubits \cite{whittaker2014} and some trapped-ion qubit encodings \cite{myerson2008}.
Within the architecture of this work, alternate mapping operations (e.g., ``latching'' \cite{studenikin2012,mason2015,harveycollard2018} and ``avalanche'' \cite{van2020} mechanisms) can provide more favorable SNR and/or relaxation rates, but potentially at added cost in mapping error.  
In this work mapping errors are principally caused by dephasing due to magnetic noise and with Landau-Zener processes, and to a lesser extent relaxation and the risk of transiting the boundaries of other charge cells.
We are also at risk of state degradation during the settling period required before source-drain biases are applied, which we believe is constrained by signal integrity limitations.
We budget some of this error by comparing two experiments: one where we bias directly from initialization to measurement, and a second where we bias from initialization to the idle bias before measurement.
These two measurements yield triplet fractions of 6\expn{-4} and 8\expn{-4}, respectively, with the difference reflecting mapping error due to that additional bias trajectory.

While mapping errors could be reduced by engineering devices with, e.g. longer singlet-triplet dephasing time $T_2^*$ or better signal integrity, they can also be limited somewhat by careful pulse design. 
Landau-Zener transitions occur when gate voltages jump between SPAM and idle configurations too quickly, such that the ground state cannot adiabatically move between the (2,0) to (1,1) charge configurations. 
Such transitions are most likely when the charge states are near an anti-crossing, as they are during initialization and measurement, and as such can be suppressed by ramping slowly to and from the SPAM configurations. 
On the other hand, magnetic dephasing is in some ways a symptom of ramping \emph{too} slowly. 
Near the charge boundary, the same large singlet-triplet energy splitting that enables SPAM suppresses dephasing between the logical states. 
Close to idle, the energy splitting is much smaller, and dephasing is more significant. 
Consequently, long-duration ramps deep within the (1,1) charge cell unnecessarily accumulate dephasing error.
We employ a two-step approach to ramping (as noted in \refsec{section:init}) to minimize both types of error.

\end{document}